\documentclass[a4paper,11pt]{article}
\pdfoutput=1 
\usepackage{jcappub} 
\usepackage[T1]{fontenc}

\usepackage{graphics}
\usepackage{rotating}
\usepackage{subfigure}
\usepackage{pstricks}
\usepackage{color}
\usepackage{amsfonts}
\usepackage{mathrsfs}
\usepackage{epsfig}
\usepackage{amsmath,amssymb,amsthm,graphicx,latexsym,braket}
\usepackage{ulem}

\newcommand{\be}{\begin{equation}}
\newcommand{\ee}{\end{equation}}
\newcommand{\bea}{\begin{eqnarray}}
\newcommand{\eea}{\end{eqnarray}}
\newcommand{\bml}{\begin{subequations}}
\newcommand{\eml}{\end{subequations}}
\newcommand{\bfig}{\begin{figure}}
\newcommand{\efig}{\end{figure}}

\title{\textsc{\fontsize{55}{90}\selectfont \sffamily \bfseries A Proposal for Constraining Initial Vacuum by CMB}}
\author[1]{Debabrata Chandra\note{Electronic address: {deb.iitdelhi@gmail.com}}}
\author[2,3]{{ and Supratik Pal}\note{Electronic address: {supratik@isical.ac.in}}\note{Work partially done: Max-Planck-Institut f\"{u}r Astrophysik,  Karl-Schwarzschild-Str. 1, D-85741 Garching, Germany}}

\affiliation{\textit{ Physics and Applied Mathematics Unit, Indian Statistical Institute, \\203 B.T. Road, Kolkata 700 108, India}}

\abstract{We propose a theoretical framework that can possibly constrain the initial vacuum by
Cosmic Microwave Background (CMB) observations.
With a generic vacuum without any particular choice   a priori, thereby
keeping both the Bogolyubov coefficients  in the analysis,
we  compute observable parameters from two- and three-point correlation functions.
We are thus left with constraining four model parameters from the two complex Bogolyubov coefficients.
 We also
 demonstrate a method of finding out the 
constraint relations between the Bogolyubov coefficients 
using the theoretical  normalization
condition and observational data of power spectrum and bispectrum from CMB.
Finally, we discuss the possible pros and cons of the analysis.}

\begin{document}
\maketitle
\flushbottom

\section{Introduction}\label{intro}

Inflationary paradigm is by now a widely accepted proposal that explains  the early universe quite meticulously as well as the formation of seeds of classical perturbations. 
In this paradigm, the different modes of quantum fluctuations of the initial state appear as classical perturbations after the horizon crossing and are measured as different observable parameters in Cosmic Microwave Background (CMB) observations\cite{wmap9,xx,xvii,xiii}. 
The CMB temperature anisotropies are the effect of the initial quantum fluctuations, hence from the CMB data one should in principle predict the nature of the primordial quantum fluctuations. The exact nature of this initial state which plays a pivotal role in providing the seeds of classical perturbations and hence observable parameters, is still eluding us.
The widely accepted initial condition is the Bunch-Davies vacuum  \cite{bunch} (BD henceforth)
analogous to Minkowski
vacuum which is strictly valid only for field theory in flat space.
It has been shown in some earlier literature that for curved spacetime one will have
 infinite number of minimum energy states \cite{daniel,bunch} and the field is in principle allowed
to choose any one of them as its initial state depending upon the physical situation. 
Yet, there are several other motivations of considering non-Bunch-Davies (NBD henceforth)
vacuum as initial condition. Some, such examples can be found in \cite{paddy,dent,chen2,meerburg,danielsson,shiu,Gasperini,
Bhattacharya,KBhattacharya,Agullo,Giovannini}.
For example, small periodic
   features in an inflation model can perturb the inflaton away from its BD state \cite{
chen2}. A preinflationary fluid dominated epoch can perturb the initial state from a zero particle vacuum to a vacuum having thermal distribution \cite{Gasperini,Bhattacharya,KBhattacharya,Agullo,Giovannini}. Also, the short distance modification of inflation, i.e., the effects of high energy
 cutoff can affect the initial condition \cite{danielsson}. 
Multifield dynamics can create excited state \cite{shiu}. In the paper \cite{meerburg} considering an effective 
field theoretic motivation they gave a weak bound too on $|\beta| < 10^{-2}$, where $\beta$ is the Bogolyubov coefficients. 
Some earlier, interesting attempts of constraining initial vacuum using different initial conditions and power spectrum alone can be found in \cite{paddy}.
We refrain ourselves from focusing on any particular mechanism or particular initial condition as such. We would rather try to bring the attention of the reader
to the fact that
 from theoretical point of view, both BD and NBD vacua have sufficient motivations to invoke.
 
CMB data, both WMAP and Planck, have helped us measure
the power spectrum and its scale dependence quite accurately but the uncertainties associated to
the measurement of the observable parameter from three-point correlation function
are huge even from the latest dataset of Planck 2015 \cite{xx,xvii,xiii}
and includes the null value as well.
A huge amount of work has been done 
  on non-Gaussian predictions for specific models of inflation\cite{
komatsu,creminelli,allen,bartolo,komaya,hamed,hall, maldacena, shell, lim, wands1, wands2, sherry, esther, zalda, paolo, 
lyth, riotto, felice, shun, kobayashi, ranu,
scsp}.
Although
   present observations suggest that the distribution is largely Gaussian,
   and we are yet to find out any considerable departure from Gaussian nature of 
   the primordial fluctuations, still primordial non-Gaussianity, even though of small amount,  are crucial 
   parameters
   to search for. 

In this article we do a generic perturbation calculation without considering
any particular  vacuum,
thereby keeping both $\alpha$ and $\beta$ in the analysis and hence without considering 
Bunch-Davies vacuum a priori.
We would rather leave it to the observational data to comment on them 
and try to constrain the allowed parameter space for those parameters from 
the available data Planck 2015. This will in turn constrain the choice of vacuum
in the theory of inflation. 
Secondly, they are in principle complex quantities, which means
there are actually four arbitrary parameters that need to be determined. We are provided with four different pieces of information, including theoretical and observational,
to determine them completely. 
Background theory and CMB data together provide three such information: (i) the normalization condition,
(ii) the power spectrum and (iii) the Bispectrum, that too  with huge errors for the third one.
Thereby, we shall engage in calculating   
 the two- and three-point correlation functions and the observables
therefrom for a generic vacuum in a model-independent way.
So, in this article, we will concentrate on the calculation of those
parameters for a generic vacuum, therefrom, constraining the parameters space of initial vacuum using those information.

Some discussions on  the present work are in order.
There are a handful of articles in the literature that explore non-Gaussianities for non-Bunch-Davies vacuum.
Some such interesting works can be found in, say \cite{kundu, gong, sandipan, burgess, agarwal, tolley, ganc1, ganc2}. 
While those works are indeed intriguing and they form the background of our work, the  objective of the present article is 
rather different. Precisely, we have two-fold goals: first,  to show a calculation of non-Gaussianities for
non-Bunch-Davies vacuum, which is as generic as possible, and secondly, to explore how far we can 
constrain the parameter space of the initial state   
directly from CMB observations.
Most of the previous articles dealing with  non-Gaussianities for non-Bunch-Davies vacuum have
either some or all of the following characteristics: they
(i)  consider only selective terms of the third order action,\cite{maldacena} (ii) consider only de Sitter solution for Mukhanov-Sasaki equation, 
(iii) calculate results by putting particular limits, e.g., squeezed limit, a priori (iv) attempt to constrain parameters from back-reaction which is not an observable. 
Our calculations are more generic in the sense that we (i) taking to account contributions from  {\it all} terms of the third order action \cite{maldacena} that makes the analysis self-consistent(which is crucial if one attempts to constrain the parameters from observation), (ii) use quasi-de Sitter solutions of Mukhanov-Sasaki equation, which is  more
accurate than usually employed de Sitter solution, so far as latest observation is concerned (iii) calculate non-Gaussian parameters (Bispectrum and $f_{NL}$) for a generic vacuum and with all possible k's, without confining 
ourselves to any particular shape a priori (we finally put different shapes only to confront with observations) (iv) attempt to constrain 
parameters directly from observations, which is more appealing.
It should be mentioned here that strictly speaking, $\alpha$ and $\beta$ are scale dependant and, in principle, one has to
take into account their scale dependance in calculations. However, as argues in some of the previous articles, 
(say, for example, \cite{shamit}),  one does not expect 
much deviation from Bunch-Davies. Hence, instead of making the already complicated calculations more messy, 
 $\alpha$ and $\beta$ can roughly be considered as constants (they are, in fact, their average values), 
without major conflict with current observations. 
Since  in this article, 
we were mostly bothered about giving a rough constraint to the parameters from present observations, we have safely 
considered them as constants.

The plan of the paper is as follows: In Section 2, we will do a brief review of the mode function calculations from second 
order perturbation of the action . Section 3 is mostly dedicated to the third order perturbation and calculation of
bispectrun therefrom for a generic vacuum keeping both $\alpha$ and $\beta$ terms. In the next two section, we show
the corresponding expressions for the variance in bispectrum, i.e. $f_{NL}$ for a  generic vacuum followed by 
its values for different limiting cases. In Section 5, we compare our analytical results with the latest observable data
from Planck 2015 and find out the possible bounds on the parameters, thereby putting constraints on the initial vacuum.
We end up with a  summary and possible future directions.
The details of the calculations using Green's function method are given
in Appendix A(\ref{appA}) and B(\ref{appB}).
%%%%%%%%%%%%%%%%%%%%%%%%%%%%%%%%%%%%%%%%%%%%%%%%%%%%%%%%%%%%%%%%%%%%%%%%%%%%%%%%%%%%%%%%%%%%%%%%%%%%%%%%%%%%%%
\section{Review of Second Order Perturbation and Power Spectrum}

For completeness, let us do a brief review of the inflationary dynamics and mode function calculations
for second order perturbation
consistent with the notations we use in this article. Some of the results from this section will also help us in 
doing numerical analysis in Section \ref{sec5}.

In what follows we will mostly concentrate on the
 Einstein-Hilbert action with canonically normalized scalar 
field minimally coupled to gravity and an arbitrary  potential \cite{daniel,maldacena,chen,liddle,langlois,collins}
\be \label{bg}
 S = \frac{1}{2}\int d^{4}x\sqrt{-g}\;\left[\;M^{2}_{pl}R\:-\left(\bigtriangledown\phi\right)^{2} -2V\left(\phi\right)\;\right]
 \ee
where $\quad M_{pl}\;=\; (8\pi \;G)^{-\frac{1}{2}}$ , $R$ is Ricci curvature, $\phi$ is the inflaton field, and $V(\phi)$ 
the inflaton potential. 
Barring some non-canonical field(s) and non-minimal coupling, this action is fairly general and can represent a
 wide class of inflationary models
\cite{baumann,daniel,maldacena,chen,liddle,guth,linde,collins}.

The background metric is as usual the FRW metric but with a somewhat unusual notation \cite{maldacena}
 \be
 ds^{2}=-dt^{2}+e^{2\rho\left(t\right)}\delta_{ij} dx^{i}dx^{j}
 \ee
 where $e^{\rho(t)}$ is the scale factor. This unusual notation for the scale factor will help the
  complicated expressions 
 (for the parameters derived from perturbative action) look
 a bit simplified. In this notation, the Hubble parameter just looks like
 \be
 \textit{H}\left(t\right) = \dot{\rho}\left(t\right) \ee
 and the background Friedmann equations and Klein-Gordon equation turn out to be
\begin{eqnarray}3M^{2}_{pl}\dot{\rho}^{2}= \frac{1}{2} \dot{\phi}^{2}+V \\
\ddot{\rho}=-\frac{1}{2M^{2}_{pl}}\dot{\phi}^{2}\\
\ddot{\phi} +  3\dot{\rho}\dot{\phi}+\frac{\partial V}{\partial \phi}=0\end{eqnarray}
 Consequently, the slow roll parameters take the form
\begin{eqnarray}
\epsilon=-\frac{\ddot{\rho}}{\dot{\rho}^{2}} = \frac{1}{2}\frac{1}{M^{2}_{pl}} 
\frac{\dot{\phi}^{2}}{\dot{\rho}^{2}} ~~~ ~
\qquad\qquad 
\eta =-\frac{\ddot{\phi}}{\dot{\rho}\dot{\phi}}\end{eqnarray}
We will make frequent use of this form of $\epsilon$.

The order-by-order perturbation of the background action (\ref{bg}) gives rise to the observable parameters.
The primordial perturbation techniques for inflationary models are well discussed in 
\cite{maldacena} and reviewed in \cite{baumann,daniel,mukhanov,collins,chen}
and are very well known in the community. 
As mentioned, we will do a brief review of the mode function calculations for the sake of subsequent sections.

Parameterizing the scalar fluctuations in terms of the gauge invariant variable  $\zeta$ \cite{bardeen}, 
considering the comoving gauge in which there is no fluctuation in the scalar field, $\delta\phi =$ 0 
and expanding the action (\ref{bg}) to second order gives \cite{maldacena}
\be 
S^{(2)}=\frac{1}{2} \int d^{4} x \left(e^{3\rho} \frac{\dot{\phi}^{2}}{\dot{\rho}^{2}} \dot{\zeta}^{2} -e^{\rho} 
\frac{\dot{\phi}^{2}}{\dot{\rho}^{2}}\partial_{k} \zeta  \partial^{k}\zeta \right)\ee

As is well known, the above action can be recast in terms of 
the Mukhanov variable
$ v=z\zeta$ (where $z=e^{\rho}\frac{\dot{\phi}}{\dot{\rho}}$) as
\be \label{S2}
 S^{(2)}=\frac{1}{2}\int d \tau d^{3} \textbf{x} \left[\left(v'\right)^{2}-\left(\partial_{i} v\right)^{2}+\frac{{z}''}{z}v \right]\ee
where  primes denote derivatives with respect to conformal time $\tau$.
Quantizing the Mukhanov variable $v$
\be  \hat{v}\left(\tau, \textbf{x}\right)= \int \frac{d^3 \textbf{k}}{(2 \pi)^3} \left[v_{k}\left(\tau\right) \hat{a}_{\textbf{k}} 
e^{i\textbf{k}.\textbf{x}}+v^{*}_{k}\left(\tau\right) \hat{a}^{+}_{\textbf{k}} e^{-i\textbf{k}.\textbf{x}}\right]\ee
and varying the second order action (\ref{S2}) one arrives at the Mukhanov-Sasaki equation 
\be {v}{}''_{k}\left(\tau\right)+\left(k^{2}-\frac{z{}''}{z} \right)v_{k}\left(\tau\right)=0 \ee
where
\be\label{mukhva} \frac{{z}''}{z}=\frac{\nu^{2}-\frac{1}{4}}{\tau^{2}}. \ee

For perfect de Sitter universe the parameter $\nu = 3/2$. But recent observations from Planck 2015 data confirms a spectral tilt ($n_s \neq$ 1) at 5$\sigma$. So, stricly speaking, one can no longer put $\nu = 3/2$ in the above equation (\ref{mukhva}) and in subsequent calculations. Rather, one needs to consider a value for the parameter $\nu$ consistent with latest value for scalar spectral index ($n_s$) as obtained from Planck 2015, say.

This equation has an exact solution in terms of Hankel function of 1st and 2nd kind\cite{baumann,dent,chen,amjad}
\be \label{mode}
v_{k}\left(\tau\right)=\sqrt{-\tau}\left[\alpha H^{(1)}_{\nu}\left(-k\tau\right)+\beta H^{(2)}_{\nu}\left(-k\tau\right)\right]
\ee
Here $\alpha$ and $\beta$ are otherwise arbitrary constants (Bogolyubov coefficients)
the values of which are determined from the initial conditions.
From the Wronskian condition for the mode function $v_{k}(\tau)$ we have a relation between $\alpha$ and $\beta$
\be\label{norm}
\left |\alpha\right|^2 - \left |\beta\right|^2 = 1 
\ee 
One should note that the above condition is a generic one which must be satisfied irrespective of the choice of vacuum.
We will use the above relation later on as the first, and generic, information to evaluate $\alpha$ and $\beta$.

A brief discussion on $\alpha$ and $\beta$ is in order. The widely accepted initial condition is Bunch-Davies (BD) initial condition
with $\alpha =1$ and $\beta=0$. This initial condition is analogous to Minkowski
vacuum which is strictly valid only for field theory in flat space.
For any other  choice of the vacuum apart from BD, (NBD henceforth) $\alpha$ and $\beta$ take 
more complicated values. 
A generic calculation for two point correlation function from the above second order perturbations using generic vacuum is already there in the literature. Below is the brief review of the same.

The power spectrum of the comoving curvature perturbation $\zeta$ can now be readily calculated from 
\be P_{k}=\left |\zeta\left(k,\tau\right)\right|^{2}=\frac{1}{z^{2}}\left |v_{k}\left(\tau\right)\right |^{2}\ee
Using the mode function equation (\ref{mode}) considering both $\alpha$ and $\beta$ one gets,
\be P_{k}=\frac{\left(-\tau\right)}{z^{2}}\left[\left(\left|\alpha\right|^{2}+|\beta|^{2}\right)\left|H^{(1)}_{\nu}\right|^{2}+2Re(\beta^{*}\alpha)\left(H^{(1)}_{\nu}\right)^{2}\right]\ee
Defining dimensionless power spectrum
 \be
 \triangle^2_s = \frac{k^3}{2 \pi^2} P_k
 \ee
 and using the boundary condition for Hankel function  
 \[\lim_{k\tau\to 0} H^{(1)}_{\nu} = \frac{i}{\pi} \Gamma(\nu)\left(-\frac{k\tau}{2}\right)^{-\nu}\]
 which is applicable for the superhorizon scales when the modes are well outside the horizon,
 the dimensionless power spectrum for super horizon limit $|k\tau |\ll 1$ takes the form
\be \label{ps}
\triangle^2_s  = \frac{\left[\Gamma(\nu)\right]^2 H^2 \left(1+\epsilon\right)^{\left(1 - 2 \nu\right)}}{2^{\left(2 - 2 \nu\right)} \pi^4 M^2_{pl}\epsilon }\left(\frac{k}{aH}\right)^{\left(3-2 \nu\right)}\left[|\alpha|^2+|\beta|^2-2Re\left(\alpha \beta^* \right)\right].
\ee
It is worthwhile to keep in mind the difference from BD vacuum (for which $\alpha =1$ and $\beta=0$ a priori).

The above relation (\ref{ps}) serves as the second piece of information and we will make
 use of its latest observational value 
from Planck 2015 to constrain the initial vacuum.

%%%%%%%%%%%%%%%%%%%%%%%%%%%%%%%%%%%%%%%%%%%%%%%%%%%%%%%%%%%%%%%%%%%%%%%%%%%%%%%%%%%%%%%%%%%%%%%%%%%%%%%%%%%%%%%
\section{Calculation of Bispectrum for Generic Initial State}\label{sec3}

As obvious from the last section the power spectrum for generic vacuum and for quasi-de Sitter solution is given by equation (\ref{ps}).  
In this article, our primary intention is to carry forward these  theoretical calculations to third order perturbations without 
any particular choice of vacuum and for quasi-de Sitter mode function. Consequently we will keep both $\alpha$ and $\beta$ in our analysis of third order perturbations. 
We remind the interested reader that our calculation is more generic than the existing literature on non-Bunch-Davies vacuum for either of the four reasons pointed out in the Introduction (\ref{intro}).

We will further employ observational data to comment on the Bogolyubov coefficients
and try to give a possible constraint on those parameters from 
the available data. This will in turn constrain the choice of vacuum in the theory of inflation. 
 
\subsection{Third Order Perturbation and Interaction Hamiltonian}
 
In order to find out the bispectrum for a generic vacuum, we first need  to expand the
 action (\ref{bg}) to third order \cite{daniel,maldacena,chen,collins} 
%\frac{\dot{\phi}^4}{\dot{\rho}^4}
% \dot{\zeta}^2\zeta +\frac{1}{4} \frac{e^{\rho}}{M^2_{pl}} \frac{\dot{\phi}^4}{\dot{\rho}^4}
% \zeta \partial_k \zeta \partial^k \zeta-\frac{1}{2} \frac{e^{3\rho}}{M^2_{pl}}
% \frac{\dot{\phi}^4}{\dot{\rho}^4} \dot{\zeta} \partial_k \zeta \partial^k \left(\partial^{-2} \dot{\zeta} \right )\notag\right. \\  \left.+ &\frac{1}{2} e^{3\rho} \frac{\dot{\phi}^2 }{\dot{\rho}^2 }
 %\dot{\zeta} \zeta^2 \frac{d}{dt} \left( \frac{\ddot{\phi}}{\dot\phi \dot{\rho}} + \frac{1}{2}
 %\frac{1}{M^2_{pl}} \frac{\dot{\phi}^2}{\dot{\rho}^2}\right) - \frac{1}{16} \frac{e^{3 \rho}}{M^4_{pl}}
% \frac{\dot{\phi}^6}{\dot{\rho}^6} \left[ \dot{\zeta}^2 \zeta - \zeta \partial_k \partial_l 
% \left(\partial^{-2} \dot{\zeta}\right )\partial^k \partial^l\left( \partial^{-2} \dot{\zeta}\right )\right]\notag\right. \\  \left.+ &f(\zeta)\left[ \frac{d}{dt}\left(e^{3 \rho}\frac{\dot{\phi}^2}{\dot{\rho}^2} \dot{\zeta} 
% \right) - e^{\rho} \frac{\dot{\phi}^2}{\dot{\rho}^2} \partial_k \partial^k \zeta \right] \right \}
%\end{align} 
\begin{multline}
S^{(3)}=\int d^4x \left \{ \frac{1}{4} \frac{e^{3\rho}}{M^2_{pl}} \frac{\dot{\phi}^4}{\dot{\rho}^4}
\dot{\zeta}^2\zeta +\frac{1}{4} \frac{e^{\rho}}{M^2_{pl}} \frac{\dot{\phi}^4}{\dot{\rho}^4}
\zeta \partial_k \zeta \partial^k \zeta-\frac{1}{2} \frac{e^{3\rho}}{M^2_{pl}}
\frac{\dot{\phi}^4}{\dot{\rho}^4} \dot{\zeta} \partial_k \zeta \partial^k \left(\partial^{-2} \dot{\zeta} \right ) \right. \\  \qquad \quad \left.+\frac{1}{2} e^{3\rho} \frac{\dot{\phi}^2 }{\dot{\rho}^2 }
\dot{\zeta} \zeta^2 \frac{d}{dt} \left( \frac{\ddot{\phi}}{\dot\phi \dot{\rho}} + \frac{1}{2}
\frac{1}{M^2_{pl}} \frac{\dot{\phi}^2}{\dot{\rho}^2}\right) - \frac{1}{16} \frac{e^{3 \rho}}{M^4_{pl}}
\frac{\dot{\phi}^6}{\dot{\rho}^6} \left[ \dot{\zeta}^2 \zeta - \zeta \partial_k \partial_l 
\left(\partial^{-2} \dot{\zeta}\right )\partial^k \partial^l\left( \partial^{-2} \dot{\zeta}\right )\right]\right. \\ \left.+ f(\zeta)\left[ \frac{d}{dt}\left(e^{3 \rho}\frac{\dot{\phi}^2}{\dot{\rho}^2} \dot{\zeta} 
\right) - e^{\rho} \frac{\dot{\phi}^2}{\dot{\rho}^2} \partial_k \partial^k \zeta \right] \right \}
\end{multline}
%\[ S^{(3)}=\int d^4x \Bigg[ \frac{1}{4} \frac{e^{3\rho}}{M^2_{pl}} \frac{\dot{\phi}^4}{\dot{\rho}^4}
% \dot{\zeta}^2\zeta +\frac{1}{4} \frac{e^{\rho}}{M^2_{pl}} \frac{\dot{\phi}^4}{\dot{\rho}^4}
%  \zeta \partial_k \zeta \partial^k \zeta-\frac{1}{2} \frac{e^{3\rho}}{M^2_{pl}}
 %  \frac{\dot{\phi}^4}{\dot{\rho}^4} \dot{\zeta} \partial_k \zeta \partial^k (\partial^{-2} \dot{\zeta})\] 
%\[  \qquad \qquad \qquad + \frac{1}{2} e^{3\rho} \frac{\dot{\phi}^2 }{\dot{\rho}^2 }
% \dot{\zeta} \zeta^2 \frac{d}{dt} \Bigg[ \frac{\ddot{\phi}}{\dot\phi \dot{\rho}} + \frac{1}{2}
%  \frac{1}{M^2_{pl}} \frac{\dot{\phi}^2}{\dot{\rho}^2}\Bigg] - \frac{1}{16} \frac{e^{3 \rho}}{M^4_{pl}}
 %  \frac{\dot{\phi}^6}{\dot{\rho}^6} \Bigg[ \dot{\zeta}^2 \zeta - \zeta \partial_k \partial_l 
 %  (\partial^{-2} \dot{\zeta})\partial^k \partial^l(\partial^{-2} \dot{\zeta})\Bigg]\]
%\begin{eqnarray} + f(\zeta)\Bigg\lbrace \frac{d}{dt}\Bigg[e^{3 \rho}\frac{\dot{\phi}^2}{\dot{\rho}^2} \dot{\zeta} 
%\Bigg] - e^{\rho} \frac{\dot{\phi}^2}{\dot{\rho}^2} \partial_k \partial^k \zeta \Bigg\rbrace \Bigg]
% \qquad \qquad \qquad \quad  \end{eqnarray}
\linebreak
where the function $f(\zeta)$ has the following form
\begin{multline}\label{fzeta}
f(\zeta)= \frac{1}{2} \left( \frac{\ddot{\phi}}{\dot{\phi} \dot{\rho}} + \frac{1}{2} 
\frac{1}{M^2_{pl}} \frac{\dot{\phi}^2}{\dot{\rho}^2} \right) \zeta^2 + \frac{1}{\dot{\rho}}
  \dot{\zeta} \zeta - \frac{1}{4} \frac{e^{-2 \rho}}{\dot{\rho}^2} \left[\partial_k\zeta \partial^k\zeta
   - \partial^{-2} \partial_{k} \partial_{l}\left(\partial^{k}\zeta \partial^{l}\zeta\right) \right] \\
   + \frac{1}{4} \frac{1}{M^2_{pl}} \frac{\dot{\phi}^2}{\dot{\rho}^3} \left\{\partial_{k}\zeta \partial^{k} \left(\partial^{-2}\dot{\zeta}\right)-\partial^{-2}\partial_{k}\partial_{l}
\left[\partial^{k}\zeta\partial^{l}\left(\partial^{-2}\dot{\zeta}\right)\right]\right\}
\end{multline}

%\[ f(\zeta)= \frac{1}{2} \left( \frac{\ddot{\phi}}{\dot{\phi} \dot{\rho}} + \frac{1}{2} \frac{1}{M^2_{pl}} \frac{\dot{\phi}^2}{\dot{\rho}^2} \right) \zeta^2 + \frac{1}{\dot{\rho}}  \dot{\zeta} \zeta - \frac{1}{4} \frac{e^{(-2 \rho)}}{\dot{\rho}^2}\left[\partial_k\zeta \partial^k\zeta   - \partial^{\left(-2\right)} \partial_{k} \partial_{l}\left(\partial^{k}\zeta \partial^{l}\zeta\right) \right] \]
%\begin{eqnarray} \label{fzeta} + \frac{1}{4} \frac{1}{M^2_{pl}} \frac{\dot{\phi}^2}{\dot{\rho}^3} \left\{\partial_{k}\zeta \partial^{k} \left(\partial^{(-2)}\dot{\zeta}\right)-\partial^{(-2)}\partial_{k}\partial_{l}\left[\partial^{k}\zeta\partial^{l}\left(\partial^{(-2)}\dot{\zeta}\right)\right]\right\} \qquad \end{eqnarray}

At the very first place, the above action looks quite difficult to handle. But, 
it can be reduced to a much simplified form 
 after a field redefinition of the gauge invariant variable $\zeta$ as following
\be \label{fr}
\zeta=\zeta_{n} + f\left(\zeta_{n}\right).
\ee
With this the above third order perturbed action takes the following form
\be \label{S3}
S^{(3)} (\zeta_{n})=\int d^4 x \frac{1}{4}\frac{1}{M^2_{pl}} \frac{\dot{\phi}^4}{\dot{\rho}^4}
 \left[ e^{3 \rho} \dot{\zeta}^2_{n} \zeta_n + e^{\rho} \zeta_n \partial_k \zeta_n \partial^k
  \zeta_n - 2e^{3\rho} \dot{\zeta_n} \partial_k \zeta_n \partial^k \left(\partial^{-2}\dot{\zeta_n}\right)\right]
\ee
We are now in a position to explore the third order perturbative action (\ref{S3}) and 
calculate the three point correlation function.
In what follows we will employ the
Schwinger-Keldysh in-in formalism \cite{schwinger,baumann,chen,collins} for computation.
In this method, the interaction Hamiltonian can be readily written from the action  (\ref{S3}) as:
\be \label{int}
H_{I}(t) =   -\frac{1}{4} \frac{1}{M^2_{pl}} \frac{\dot{\phi}^4}{\dot{\rho}^4} \int d^3 \textbf{x} \left[ e^{3 \rho} \dot{\zeta}^2_{n} \zeta_n + e^{\rho} \zeta_n \partial_k \zeta_n \partial^k
  \zeta_n - 2e^{3\rho} \dot{\zeta_n} \partial_k \zeta_n \partial^k \left(\partial^{-2}\dot{\zeta_n}\right)\right]
\ee
This interaction Hamiltonian will be used to evaluate the three point function and therefrom the Bispectrum.

\subsection{Bispectrum for generic vacuum}

The leading order non-Gaussianity is given by the Fourier transform of three 
point Correlation function usually known as 
Bispectrum, which is defined as \cite{xvii,daniel}
\be \langle\zeta_\textbf{k}\zeta_\textbf{m}\zeta_\textbf{n}\rangle =B_{\zeta}
(\textbf{k},\textbf{m},\textbf{n}) = (2\pi)^3\delta(
\textbf{k}+\textbf{m}+\textbf{n}) B_{\zeta}({k},{m},{n})\ee
One need to calculate the three point correlation function first to get the Bispectrum.
The evaluation of the three point function involves following steps,

We first expanded the three point function using the redefinition (\ref{fr}) of field $ \zeta $  ,
\[\langle 0(t)\left|\zeta(t,\textbf{x})\;\zeta(t,\textbf{y})\;\zeta(t,\textbf{z})\right|0(t)\rangle \qquad
 \qquad \qquad \qquad \qquad \qquad \qquad \qquad \qquad \qquad \qquad \qquad\]
\[=\;\langle 0(t)|\zeta_{n}(t,\textbf{x})\;\zeta_{n}(t,\textbf{y})\;\zeta_{n}(t,\textbf{z})
|0(t)\rangle+\langle 0(t)|\zeta_{n}(t,\textbf{x})\;\zeta_{n}(t,\textbf{y})\;f(\zeta_{n}
(t,\textbf{z}))|0(t)\rangle\qquad \qquad\]\[\;+\langle 0(t)|\zeta_{n}(t,\textbf{x})\;
f(\zeta_{n}(t,\textbf{y}))\;\zeta_{n}(t,\textbf{z})|0(t)\rangle\;+\;\langle 0(t)
|f\left(\zeta_{n}(t,\textbf{x})\right)\;\zeta_{n}(t,\textbf{y})\;\zeta_{n}(t,\textbf{z})|0(t)\rangle \;\]$ ~~\qquad + ....$

Here the dotted term represent 
Higher Order terms  that include at least five features of the Shifted function. 
They are suppressed due to the higher order slow roll parameter effects.
Note that the three point function $\zeta_{n}(t, \textbf{x})$ would vanish entirely 
if evaluated in the initial state and
in interaction picture its value entirely generated by the evolution of state.

Thus, in terms of the interaction Hamiltonian,  the three point correlation function takes the form
\[\langle 0(t)|\zeta_{n}(t,\textbf{x})\;\zeta_{n}(t,\textbf{y})\;\zeta_{n}(t,\textbf{z})|0(t)\rangle
 \qquad \qquad \qquad \qquad \qquad \qquad \qquad \qquad \qquad \qquad \qquad \qquad\]
\[=\;\langle 0|\left(Te^{-i \int_{-\infty}^t dt' H_{I}\left(t' \right) }\right)^{\dagger}\; \zeta_{n}
\left(t,\textbf{x}\right)\;\zeta_{n}\left(t,\textbf{y}\right)\;\zeta_{n}\left(t,\textbf{z}\right)\left(Te^{-i \int_{-\infty}^t dt' H_{I}\left(t' \right) }\right)|0\rangle \qquad \]

For the rest of the terms as $f(\zeta_{n})$ are quartic in the field $\zeta_{n}$,
 they do contribute in initial state $|0 \rangle$. In this case time evolution of the
  state only contributes at a higher order in slow roll parameter.

 Now one is left with the following two tasks:
(i) inserting the interaction Hamiltonian (\ref{int}) 
in the above expression of three point correlation function and calculate the effects of each term
using Green's function method.
(ii) evaluate the terms containing $f(\zeta_{n})$, using $|0 \rangle$ as initial state.

Using the interaction Hamiltonian (\ref{int}) and the results given in {\it Appendix A} (\ref{appA}) we calculated the three point function. One needs to keep in mind that everything has 
to be calculated for a generic vacuum using the mode function (\ref{mode}). 
This turns out to be quite a daunting task, first, due to the form of the interaction 
Hamiltonian and, secondly, because of the presence of both  $\alpha$ and $\beta$ in the mode function.

However, we have been able to find out the bispectrum for the generic case.  Final result for bispectrum that reads
\be \label{bispec}
B_{\zeta}\left(\textbf{k},\textbf{m},\textbf{n} \right) = \left(2 \pi \right)^3 \delta^3 \left(\textbf{k}+\textbf{m}+\textbf{n}\right) \frac{1}{k^3 m^3 n^3} \frac{H^4}{32 \epsilon^2 M^4_{pl}}\mathcal{A} \ee
 
In this section we haven't shown the explicit expression of $ \mathcal{A} $ to maintain the simplicity and the elegancy of the article. The details of the calculations and the explicit form of $ \mathcal{A} $ are given in {\it Appendices} (\ref{appA}) (\ref{appB}).

%In the next section for the calculation of $f_{NL}$ we explicitly show the complicated expression, to keep the consistency of the article.

%%%%%%%%%%%%%%%%%%%%%%%%%%%%%%%%%%%%%%%%%%%%%%%%%%%%%%%%%%%%%%%%%%%%%%%%%%%%%

\section{Calculation of $f_{NL}$ from Bispectrum}\label{sec4}

Even though after painstaking exercise we have succeeded in finding out the bispectrum 
for a generic vacuum with quasi-de Sitter mode function, if one really wants to 
constrain the vacuum using observations, one needs to proceed further and 
calculate its variance, namely, the $f_{NL}$.
Of course, we haven't yet detected non-Gaussianities, so any value of $f_{NL}$ 
in the range predicted by Planck 2015 is allowed, we will show in the subsequent
 section how the present information alone
can help us constrain the vacuum to some extent.

Using the expression from bispectrum from Eq (\ref{bispec}), one can define its variance
$f_{NL}$ as \cite{maldacena}
\begin{equation}\label{fnl}
-f_{NL} \sim \frac{5}{3} \frac{\mathcal{A}}{\left(4 \sum_{i} k^3_i\right)}  = \frac{5}{12} 
\frac{\mathcal{A}}{\left(k^3+m^3+n^3\right)}
\end{equation}
Here one can look into the above expression of $f_{NL}$  in an explicit form using the variable $\mathcal{A}$, which has been shown in the  {\it Appendix B} (\ref{appB}) in a comprehensive way.

Below 
we will explore different shapes of mode to evaluate $f_{NL}$ which are relevant for
 observational purpose \cite{xvii} to constrain the initial state of inflationary dynamics.

\subsection{$f_{NL}$ in Squeezed Limit} 

Here we considered squeezed limit case where one of the modes is very small relative to the other two modes 
\[\textbf{k}\longrightarrow 0 \quad \textbf{m} \approx -\textbf{n}~~ or ~~ m \approx n\]
Using the generic expression of $f_{NL}$ from Eq (\ref{fnl}) and applying the above condition we get
\begin{multline}
-f^{\rm squeez}_{NL} =\frac{40  \epsilon^2 M^4_{pl}}{3 H^4} \left(\epsilon - \eta \right)A^2 + \frac{80 \epsilon^2 M^4_{pl}}{3 H^5}\left( 4 - \epsilon \right)AB + \frac{160 i \epsilon^4 M^6_{pl}}{3\left(1-\epsilon \right)^4 H^8}\left( D_3+D_4  \right)\qquad\qquad \qquad \qquad ~~~~~~~~~~~~~~~ \\  +  \left[ \frac{160 i \epsilon^4 M^6_{pl}}{3 \left(1-\epsilon \right)^2 H^6} C_3 C_4 \left(C_3 + C_4 \right) + \frac{160 i \epsilon^4 M^6_{pl}}{3 \left(1-\epsilon \right)^4 H^8} \left(D_3+D_4 \right) \right]\frac{m}{k}  ~~~~~~~~~~~~
\end{multline}
%\[-f^{\rm squeez}_{NL} = \frac{5}{12} \left[\left(\frac{\ddot{\phi}}{\dot{\phi}\dot{\rho}}+ \frac{1}{2}
% \frac{1}{M^2_{pl}} \frac{\dot{\phi}^2}{\dot{\rho}^2}\right) A^2 + \frac{2}{\dot{\rho}}AB - 
 %\frac{1}{4} \frac{1}{M^2_{pl}} \frac{\dot{\phi}^2}{\dot{\rho}^3}AB -
%  \frac{2i \epsilon^2 M^2_{pl}}{H^4 (1-\epsilon)^4} D_1 \right] \frac{32 \epsilon^2 M^4_{pl}}{H^4} \]
   %\[  + \frac{5}{12}\Bigg\lbrace\frac{128 i \epsilon^4 M^6_{pl}}{(1-\epsilon)^4 H^8}(D_3+D_4) + 
   % \frac{64 i \epsilon^4 M^6_{pl}}{(1-\epsilon)^4 H^8} D_1 \qquad \qquad \qquad \qquad  \]
% \begin{eqnarray} \qquad+ \left[ \frac{128 i \epsilon^4 M^6_{pl}}{(1-\epsilon)^4 H^8}(D_3+D_4)
 %  +  \frac{128 i \epsilon^4 M^6_{pl}}{(1-\epsilon)^2 H^6} C_3 C_4 (C_3 + C_4)\right]\frac{m}{k} \Bigg\rbrace
 %   \qquad \quad \end{eqnarray}

 Note that unlike the BD case, there is a term with $\frac{m}{k}$ sitting in the expression for $f_{NL}$ in Squeezed limit. 
 This arises solely due to a nonzero  $\beta$ and the expression cannot be 
 readily reduced to that of the BD case.
 The possible conclusions from the above expression for squeezed limit can be the following:
 
 \begin{itemize}
\item Here we have got a new result that $f_{NL}$ in squeezed limit is scale dependent 
as we can find that $f_{NL}$ is having a $\frac{m}{k}$ term. So here we may conclude that
 for non Bunch-Davies vacuum we can get scale dependent squeezed limit value of  $f_{NL}$
  for single field slow-roll model.

\item Since in the squeezed limit, $k\longrightarrow 0$, maybe future detection of $f_{NL}$ 
in squeezed limit will set a scale of $k$ and will answer how much 'squeezed' it can be.
\item In the worst case, $f_{NL}$ for squeezed limit cannot be estimated for any vacuum 
other than the Bunch-Davies one. 

\end{itemize}

However,  we must admit that all of these are mere speculations and we refrain ourselves 
from making a definite comment on this. We would rather make use of the $f_{NL}$ in
 Equilateral limit calculated in the next subsection for
constraining the initial vacuum.

\subsection{$f_{NL}$ in Equilateral Limit} 

For another shape of modes, where all modes are equal in magnitude known as
 equilateral shape we calculated the $f_{NL}$ 
\[k=m=n\]
Similarly as done in squeezed limit case applying the above relation of modes 
in the generic expression for $f_{NL}$ from Eq (\ref{fnl}), we find
\begin{multline}\label{eql}
-f^{\rm equil}_{NL} =  \frac{40 \epsilon^2 M_{pl}^4}{3 H^4} \left( \epsilon-\eta \right)A^2 + \frac{10 \epsilon^2 M_{pl}^4}{ H^5} \left( \frac{8}{3}-\epsilon \right)AB +\frac{160 i \epsilon^4 M^6_{pl}}{3 \left(1-\epsilon \right)^2 H^6} \left[ \frac{1}{18}\left(C^3_3+C^3_4 \right) \right. \\ \left. + \frac{3}{2}C_3 C_4 \left(C_3+C_4 \right)\right] +\frac{160 i \epsilon^4 M^6_{pl}}{3 \left(1-\epsilon \right)^4 H^8} \left[\frac{2}{9}D_1 + 2 D_2 + 4 \left(D_3 + D_4 \right)\right] 	~  \qquad \quad\end{multline}
%\[-f^{\rm equil}_{NL} = \frac{5}{12} \left[\left(\frac{\ddot{\phi}}{\dot{\phi}\dot{\rho}}
 %+ \frac{1}{2} \frac{1}{M^2_{pl}} \frac{\dot{\phi}^2}{\dot{\rho}^2}\right)A^2 +
%  \frac{2}{\dot{\rho}}AB - \frac{1}{4} \frac{1}{M^2_{pl}} \frac{\dot{\phi}^2}
%  {\dot{\rho}^3}AB -\frac{2i\epsilon^2 M^2_{pl}}{H^4(1-\epsilon)^4}D_1\right]\frac{32 \epsilon^2 M^4_{pl}}{H^4}\]

%\[~~~~~ \qquad\qquad\qquad+ \frac{5}{12}\Bigg\lbrace-\frac{1}{8} \frac{1}{M^2_{pl}} 
%\frac{\dot{\phi}^2}{\dot{\rho}^3} \frac{32 \epsilon^2 M^4_{pl}}{H^4}AB  + 
%\frac{128 i \epsilon^4 M^6_{pl}}{(1-\epsilon)^4 H^8}\Bigg[ \frac{D_1}{9} +
%  D_2 +2(D_3+D_4)\Bigg] \qquad \qquad \qquad \qquad\]

%\[~~ + \frac{128 i \epsilon^4 M^6_{pl}}{(1-\epsilon)^2 H^6} \Bigg[ \frac{1}{18}(C^3_3+C^3_4)
% + \frac{3}{2}C_3 C_4(C_3+C_4)\Bigg] 
%\qquad \qquad \qquad\]
%\be\label{eql} +\frac{128 i \epsilon^4 M^6_{pl}}{(1-\epsilon)^4 H^8} 
%\Bigg[\frac{11}{18}D_1 + D_2 + 2(D_3 + D_4)\Bigg]\Bigg\rbrace 
%\qquad \qquad \qquad \qquad \ee 
 where, as before, the quantities $A$, $B$, $C_i$ and $D_i$ contain all the
  information about a generic vacuum and hence about the parameters $\alpha$ and $\beta$.
 
$f_{NL}$ in Equilateral limit in Eq (\ref{eql}) will serve as the complimentary
 piece of information along with 
the normalization condition Eq (\ref{norm}) and power spectrum Eq (\ref{ps}) to
 constrain the allowed parameter space for 
initial vacuum. In what follows we shall explore this in details.

%%%%%%%%%%%%%%%%%%%%%%%%%%%%%%%%%%%%%%%%%%%%%%%%%%%%%%%%%%%%%%%%%%%

\section{Possible Constraints on The Parameters from Planck 2015}\label{sec5} 

As pointed out earlier, for a generic vacuum, the parameters $\alpha$ and $\beta$  are
 in principle complex quantities, which means
there are actually four arbitrary parameters that need to be determined from initial
 conditions. So in principle,
one needs to provide  four different pieces of information, either from theoretical 
consideration or from observational constraints,
to determine them completely. In this article we have explored this possibility and
 found that we can have three pieces of information
from early universe and CMB:

\begin{itemize}
 \item the normalization condition Eq (\ref{norm}) which is generic one and must
  be satisfied for any vacuum, irrespective of whether it is BD or NBD.
 \item the power spectrum for a generic vacuum (\ref{ps}) and its numerical value 
 from Planck 2015
 \item information from the bispectrum, in particular, the $f_{NL}$ in Equilateral
  limit in Eq (\ref{eql}) and its bound as given by Planck 2015.
\end{itemize}

Given that we are going to determine four unknown parameters but we are given only
 three pieces of information, 
we are thus left with the choice of finding out the allowed parameter space  for
 $\alpha$ and $\beta$.
On top of that, given that we have not yet detected primordial non-Gaussianities
 and Planck 2015 can at best give some bounds on 
$f_{NL}$, we have to deal with the uncertainty of having zero non-Gaussianity as
 well, thereby losing this third piece of information.
We would rather  be somewhat optimistic and take the bounds on $f_{NL}$ at the 
face value.
As it appears, this will indeed help us in constraining the allowed parameter 
space  for $\alpha$ and $\beta$
to some extent which will in turn put constraints on the choice of the vacuum.
\subsection{Parameter Spaces from Planck 2015}

Below we will find out the constrain relations between  $\alpha$ and $\beta$ 
from the previously mentioned information.
The first constraint relation is straightforward and it is given by 
the normalization condition (\ref{norm}):
\be\label{const1}
|\alpha|^2 - |\beta|^2 = 1 
\ee 

From Eq (\ref{ps}), the  amplitude of power spectrum for this generic vacuum is given by
\be\label{a_s}
A_s = \frac{ \left[\Gamma \left(\nu \right) \right]^2 H^2 \left(1+\epsilon\right)^{\left(1 - 2 \nu\right)}}{2^{\left(2 - 2 \nu \right)} \pi^4 M^2_{pl} \epsilon }\left[|\alpha|^2+|\beta|^2-2Re \left( \beta^* \alpha \right)\right]\qquad \quad \ee

As we have mentioned earlier for perfect de-Sitter universe the parameter $\nu = 3/2$. But recent observations from Planck 2015 data confirms a spectral tilt ($n_s \neq$ 1) at 5$\sigma$. So, strictly speaking, one can no longer put $\nu = 3/2$ in the expressions for power spectrum, bispectrum and $f_{NL}$, as done in most of the existing literature. Rather, one needs to consider a value for the parameter $\nu$ consistent with latest value for scalar spectral index ($n_s$) as obtained from Planck 2015. This is precisely what we are going to do in the subsequent calculations. Our analysis is, therefore, more up-to-date and consistent with latest observations.

The value of  scalar spectral index ($n_s$) as obtained from Planck 2015 is $n_s = 0.9677 \pm 0.0060~~ $(TT+lowP+lensing)\cite{xx}. In what follows, we will make use of the best fit value only. Using the relation between $n_s$ and $\nu$, $\nu = (2-\frac{n_s}{•2})$ we obtain the best fit value of the parameter $\nu= 1.5162$ from the best fit value of $n_s$, which we are going to use throughout.

Further, combined Planck TT + low p results give 
the best fit value for $\epsilon = 0.0068 \;[\;95\% $ CL] \cite{xx}.
Combining all these, the amplitude of scalar power spectrum for a generic vacuum 
turns out to be
\be
A_s = {2.8241}\times 10^{-9}\left[|\alpha|^2 + |\beta|^2 - 2 Re\left(\beta^*\alpha\right)\right]
\ee \label{amp}

One can now compare this result with the Planck 2015 data  \cite{xx}. From combined 
Planck TT + low p + lensing, we have
$\ln (10^{10} A_s) = 3.062 \pm 0.029\ $ [$\;68\%$ CL]
Strictly speaking, one should consider the best fit value along with the error bars
 in the analysis, exactly what we have done in the subsequent section of rigorous approach. However, for this present analysis we can safely neglect the error in power spectrum and take into account its 
best fit value only because of the fact that 
the error bars in power spectrum are  small compared to that of bispectrum 
(which we will make use of in the next step). This will somewhat simplify our analysis.
Using the best fit value, Eq (5.3) give rise to a constraint relation
 between $\alpha$ and $\beta$ as 
\be\label{const2}
|\alpha|^2 + |\beta|^2 - 2Re\left(\beta^*\alpha\right) = 0.7567
\ee 

The third constraint relation comes from exploring the $f_{NL}$ in
 Equilateral limit.
Expressing explicitly in terms of $\alpha$ and $\beta$ via the parameters
 $A$, $B$, $C_i$, $D_i$ etc, 
and using the best fit value for  $\eta_p = 0.029\;[\;68\;\%$  CL Planck
 TT + low p ] \cite{xx} 
(where $- \eta = \frac{\eta_p}{2} - \epsilon$), along with the previously
 used values
for $\nu$ and $\epsilon$,
the $f^{equil}_{NL}$ takes the following relatively simple form 
\begin{multline}
-f^{equil}_{NL} = 0.0429\left(|\alpha|^2 - \alpha^* \beta\right)
\left(|\beta|^2 - \beta^*\alpha\right)\left(|\alpha|^2 - |\beta|^2 - \alpha^* \beta + \beta^* \alpha\right) \\ -~0.0184 \left(|\alpha|^2 + |\beta|^2 - \alpha^* \beta - \beta^*\alpha\right)^2
\end{multline}
%\[-f^{equil}_{NL} = \frac{5}{12}\times\left(0.0049\right)\Bigg[\left(21.0143\right)\left(|\alpha|^2 - \alpha^* \beta\right)
%\left(|\beta|^2 - \beta^*\alpha\right)\left(|\alpha|^2 - |\beta|^2 - \alpha^* \beta + \beta^* \alpha\right)\]
%\begin{eqnarray} - \left(9.0225\right)\left(|\alpha|^2 + |\beta|^2 - \alpha^* \beta - \beta^*\alpha\right)^2\Bigg]
% \qquad \qquad \qquad \qquad \quad \quad \end{eqnarray}

As is well known, the 
observational bounds for $f_{NL}$ comes with huge errors. Also, the uncertainties
include the zero non-Gaussinities as well. 
Of course, future observations can improve the scenario further and in principle,
 one can use the improved value of $f_{NL}$
to constrain the vacuum following our analysis.
At this moment, the best one can do is to take into account the huge errors in the
 analysis and search for the  maximum 
possible information till date. 

From Planck 2015 \cite{xvii}[$\;68\%$ CL temp\; and\; polarization]
we have
\be
f^{equil}_{NL} = -4\pm43
\ee 
which, when compared to the above expression, gives the third constraint relation
 between $\alpha$ and $\beta$ as below
\begin{multline}\label{const3}
-0.0429\left(|\alpha|^2 - \alpha^*\beta\right)\left(|\beta|^2 - \beta^* \alpha\right)\left(|\alpha|^2 - 
|\beta|^2 - \alpha^*\beta + \beta^*\alpha\right) \\+ 0.0184\left(|\alpha|^2 + |\beta|^2 - \alpha^*\beta - \beta^* \alpha\right)^2 = -4\pm43
\end{multline}
%\[21.0143\left(|\alpha|^2 - \alpha^*\beta\right)\left(|\beta|^2 - \beta^* \alpha\right)\left(|\alpha|^2 - 
%|\beta|^2 - \alpha^*\beta + \beta^*\alpha\right) 
%\qquad \qquad \qquad \qquad \qquad\] 
%\be\label{const3}
%- 9.0225\left(|\alpha|^2 + |\beta|^2 - \alpha^*\beta - \beta^* \alpha\right)^2 = 
%-490.7976\times\left(-4\pm43\right) \qquad \qquad 
%\ee 

As pointed out earlier, we are now left with these three constraint relations
 Eq (\ref{const1}), (\ref{const2}),  (\ref{const3})
and four unknowns.
As we committed before now we will try to extract the maximum information about 
$\alpha$ and $\beta$ from the above discussed three relations.
For that purpose let us break the $\alpha$ and $\beta$ in real and imaginary parts by considering
\be
\alpha = a_1 + ia_2 \quad \quad \beta = b_1 + ib_2 \ee 
where $a_1$, $a_2$, $b_1$, $b_2$ are the four unknowns that need to be constrained.
Putting them back into the three constraint relations and subsequently, solving them numerically,
we finally arrive at the  allowed region of the parameters from Planck 2015. 

The following figures (Fig 1-3) represent  the allowed region for each pair of parameters.
For each figure,  we have chosen two parameters from the set $\{a_1, a_2, b_1, b_2 \}$ and expressed them  in terms of 
the other two. We have finally utilized the three constraint relations to find out the allowed regions
for each pair
numerically.
\begin{figure}[h] \label{fig1}
\center
\includegraphics[scale=0.50]{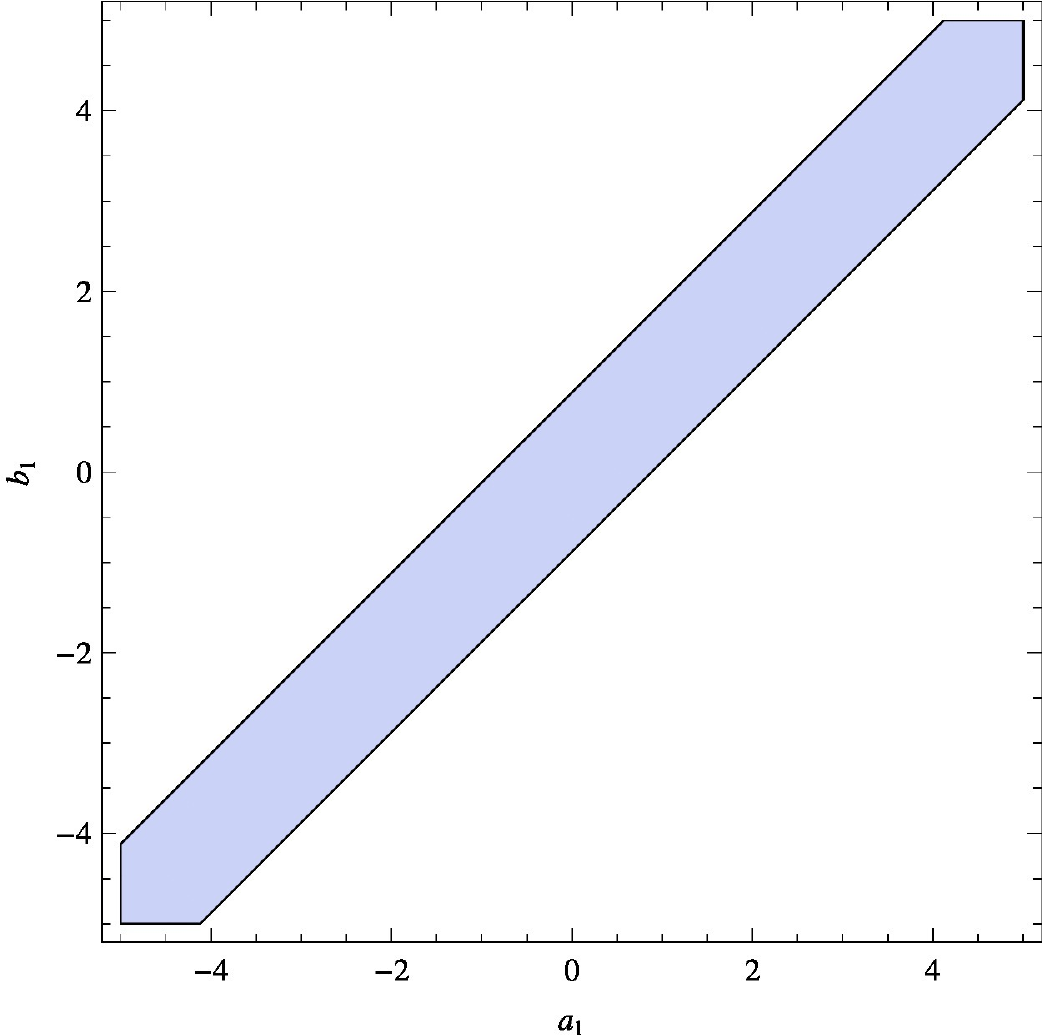}
\caption{The figure shows the allowed region for $a_1$ vs $b_1$ plot. The Bunch-Davies (BD)vacuum, which is just a point \{1, 0\} in this plot, sits comfortably inside with a spread allowing some NBD vacuua as well.}
\end{figure}

\begin{figure}[h]
\center
\includegraphics[scale=0.50]{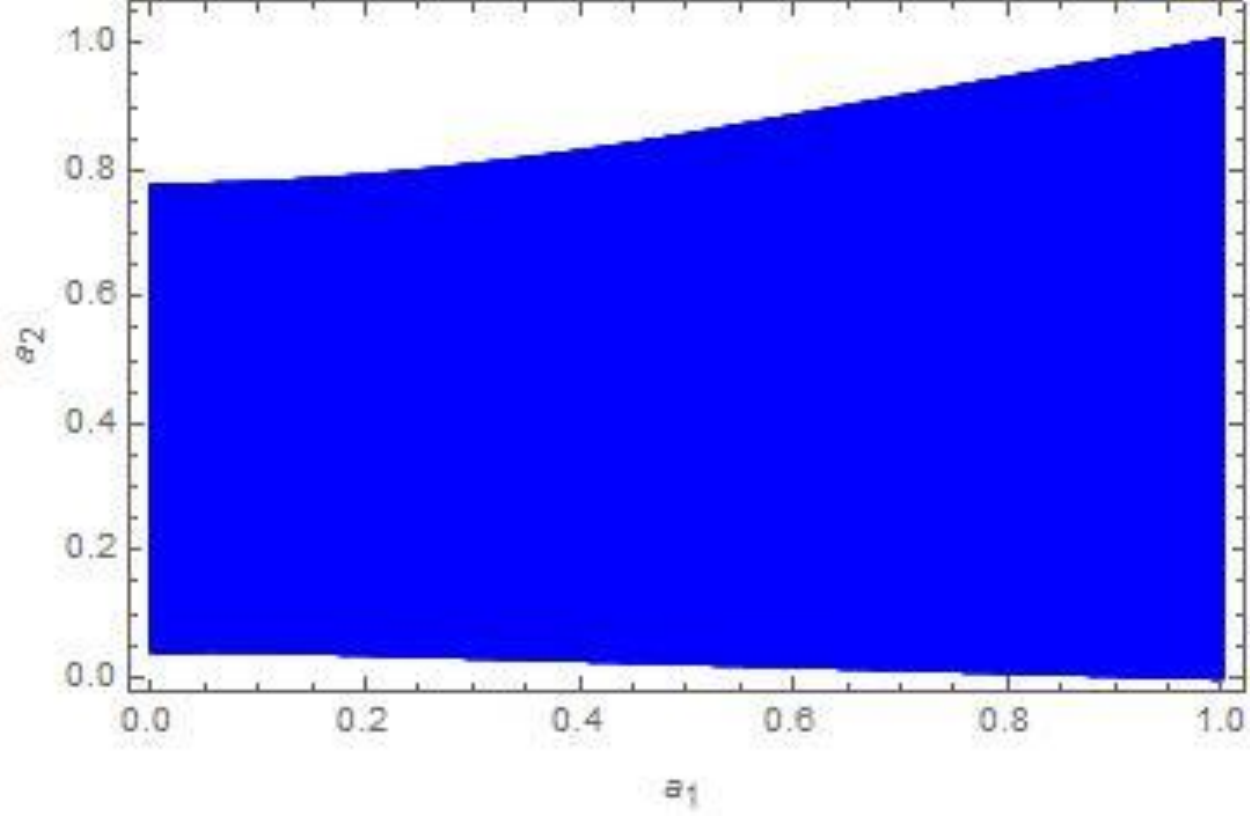}
\caption{This figure depicts the allowed region for $a_1$ vs $a_2$ considering both BD and NBD vacuum.}
\end{figure}

\begin{figure}[h]
\center
	\includegraphics[scale=0.45]{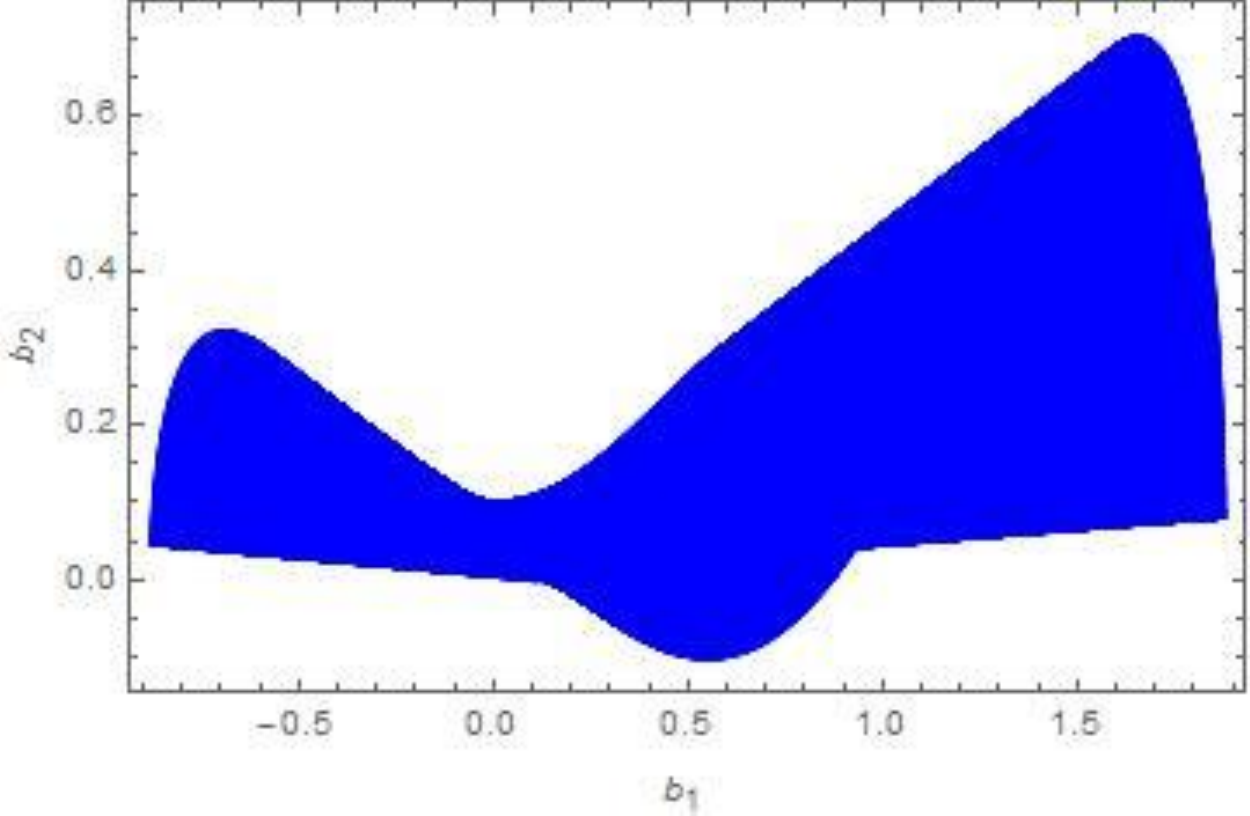}
	\caption{ This figure represents the allowed region for $b_1$ vs $b_2$ considering both BD and NBD vacuum.}
\end{figure}

%%%%%%%%%%%%%%%%%%%%%%%%%%%%%%%%%%%%%%%%%%%%%%%%%%%%%%%%%%%%%%%%%%%%%%%%%%%%%%%%
\subsection{Parameter Spaces from Planck 2015 (rigorous approach)}

In the previous section we have demonstrated the methodology of putting possible
constriants on the vacuum from CMB. This was more or less a theoretical study 
considering only the best fit values as obtained from Planck 2015.
However, most of the CMB parameters (e.g., the slow roll parameter $\epsilon$)
are obtained by assuming Bunch-Davies vacuum a priori. So, it is advisable to do
a rigorous numerical analysis without assuming any a priori values for
the Bogolyubov coefficients as well as for
the slow roll parameter $\epsilon$, along with that, using the observed error bars too, instead of considering only the best fit values. 
As apparent, this may lead to a more accurate constraint
on the initial vacuum as well as on the slow roll parameter, and may help us have 
a better understanding of the inflationary scenario from observations.
In what follows, we shall engage into this analysis. 
Specifically, we will try to find out 2-dimensional parameter spaces for different theoretical parameters
involved in the present analysis, and the best fit values of these parameters therefrom.
For this, we shall make use of the generic expressions for power spectrum and bispectrum as obtained in 
Eq (\ref{ps}) and Eq (\ref{eql}) respectively.

In order to do so, 
let us redefine the parameters by  breaking the Bogolyubov coefficients $\alpha$ and $\beta$ in real and imaginary parts by considering
\be\label{bogol}
\alpha = a e^{i \theta} \quad \quad \beta = b e^{i \phi} \ee 
where $a$, $b$, $\theta$, $\phi$ are the four unknowns that need to be constrained.

From Eq (\ref{ps}), the  amplitude of the power spectrum for this generic vacuum is given by
\[
A_s = \frac{\left[\Gamma(\nu)\right]^2 H^2 \left(1+\epsilon\right)^{\left(1 - 2 \nu\right)}}{2^{\left(2 - 2 \nu\right)} 
\pi^4 M^2_{pl} \epsilon }\left[|\alpha|^2+|\beta|^2-2Re \left( \beta^* \alpha \right)\right]\qquad \quad \]

From Eq (\ref{eql}) we have the generic expression for $f_{NL}$ in Equilateral Limit  

\begin{multline}\label{fifnl}
	-f^{equil}_{NL} =  \frac{80}{9}\frac{\epsilon \left(4\nu^2-1 \right)^3}{ \left(8\pi \right)^3} \left[\left(|\alpha|^2 -ie^{i \nu \pi}\alpha^* \beta\right)^3-\left(|\beta|^2 +ie^{-i \nu \pi}\beta^*\alpha\right)^3\right] \\
\qquad \qquad \quad ~~+ \frac{560}{3}\frac{\epsilon \left(4\nu^2-1 \right)^3}{ \left(8\pi \right)^3}\left(|\alpha|^2 -ie^{i \nu \pi}\alpha^* \beta\right)\left(|\beta|^2 +ie^{-i \nu \pi}\beta^*\alpha\right)\left(|\alpha|^2 - |\beta|^2 -ie^{i \nu \pi} \alpha^* \beta -ie^{-i \nu \pi} \beta^* \alpha\right)
\\ +\frac{5}{24}\frac{(\frac{\eta_p}{2}-4\epsilon) (4\nu^2-1)^4}{\left(8\pi \right)^2}\left(|\alpha|^2 + |\beta|^2 - ie^{i \nu \pi}\alpha^* \beta + ie^{-i \nu \pi}\beta^*\alpha\right)^2   \qquad \qquad	   
\end{multline}
%\[-f^{equil}_{NL} = \frac{5}{12}\Bigg\lbrace\frac{128\epsilon(4\nu^2-1)^3}{(8\pi)^3} \left[\frac{2}{3}[\left(|\alpha|^2 -ie^{(i \nu \pi)}\alpha^* \beta\right)^3-(|\beta|^2 +ie^{(-i \nu \pi)}\beta^*\alpha)^3\right]\]
%\[+\frac{7}{2}(|\alpha|^2 -ie^{(i \nu \pi)}\alpha^* \beta)
%(|\beta|^2 +ie^{(-i \nu \pi)}\beta^*\alpha)(|\alpha|^2 - |\beta|^2 -ie^{(i \nu \pi)} \alpha^* \beta -ie^{(-i \nu \pi)} \beta^* \alpha)\Bigg\rbrace\]
%\[+\Bigg[\frac{(4\nu^2-1)^4(\frac{\eta_p}{2}-4\epsilon)}{128\pi^2}(|\alpha|^2 + |\beta|^2 - ie^{(i \nu \pi)}\alpha^* \beta + ie^{(-i \nu \pi)}\beta^*\alpha)^2\Bigg]-\frac{\epsilon(4\nu^2-1)^3}{8\pi^3}\]
%\begin{eqnarray}\label{fifnl}\Bigg[(|\alpha|^2 -ie^{(i \nu \pi)}\alpha^* \beta)^3-
%(|\beta|^2 +ie^{(-i \nu \pi)}\beta^*\alpha)^3\Bigg]\Bigg\rbrace
% \qquad \qquad \qquad \qquad \quad \quad\end{eqnarray}
 
Let us now constrain the parameter spaces by using Eq(\ref{const1}), (\ref{a_s}), (\ref{bogol}), (\ref{fifnl}).
With the help of Planck 2015\cite{xvii,xx,xiii} data and the above equations, the parameter spaces for
$a$, $b$, $\delta\:($where $\delta=\theta-\phi)$ and $\epsilon$ have been obtained.

The following plots depict the pairwise plots for them. The first one gives possible constraints
and best fit values by considering the amplitude of the power spectrum only, whereas in the second plot 
we consider both amplitude of power spectrum and $f_{NL}^{equil}$ to constrain the respective parameter
spaces. The results are self-explanatory. As obvious, we get a more accurate constraint for the parameters from this analysis compared to the one done in the previous section.

\begin{figure}[h]
\center
	\includegraphics[scale=1.30]{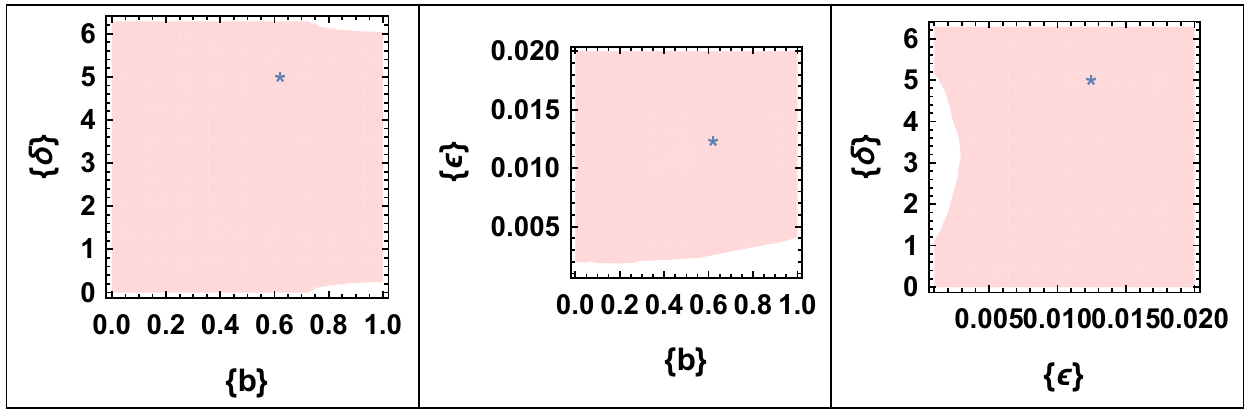}
	\caption{This figure represents the allowed region for $b$ vs $\delta$, $b$ vs $\epsilon$ and $\epsilon$ vs $\delta$, 
	considering only the analysis of $A_s$ and $`*`$ represent the best fit value $b=0.62136$, $ \delta=5.04116 $ and $ \epsilon=0.01248 $ .}
\end{figure}

\begin{figure}[h]
\center
	\includegraphics[scale=1.30]{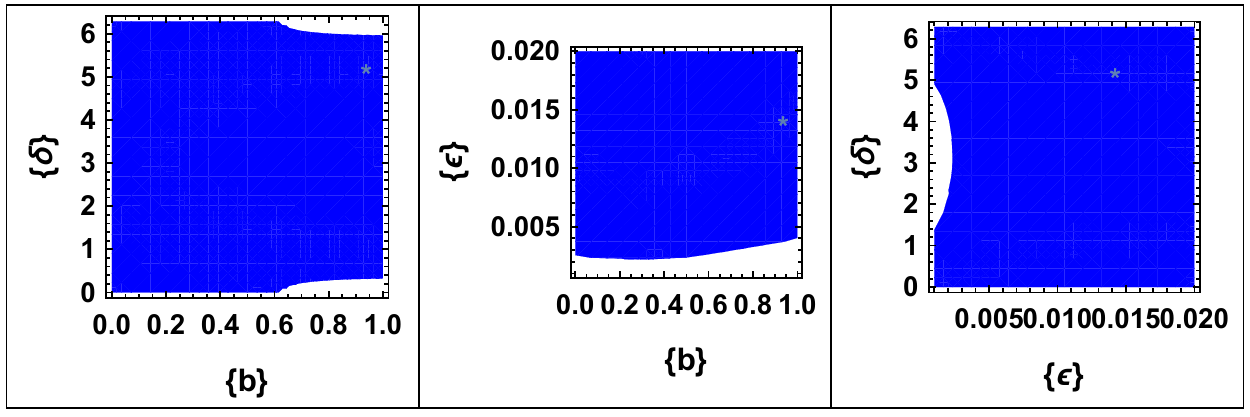}
	\caption{ This figure represents the allowed region for $b$ vs $\delta$, $b$ vs $\epsilon$ and $\epsilon$ vs $\delta$, 
	considering the joint analysis of $A_s$ and $f_{NL}^{equil}$ and $`*`$ represent the best fit 
	value $b=0.93789$, $ \delta=5.23004 $ and $ \epsilon=0.014229 $.}
\end{figure}

The above plots(Fig 1-5) summarize our results and claim. What they essentially tell us
is that the Bunch-Davies vacuum
 is still a
well-acceptable vacuum as rightly done in most of the calculations. Since, we get
 a wide allowed region in the parameter space for the Bogolyubov constants, this essentially means 
quite a handful of non-Bunch-Davies vacua
are equally allowed so far as the present data is concerned. Nevertheless, whenever 
one proposes a new
vacuum, one need to check if it falls within the allowed region of CMB observations 
following the method
demonstrated in the present article.

\section{Summary, Limitations and Future Directions}

In this article, we 
attempted to demonstrate how one can possibly constrain the initial vacuum using
 CMB.
Our analysis focuses on two major goals,
 (i) to calculate the observable parameters
for a generic vacuum (i.e., without any particular choice of the vacuum a priori), thereby
keeping both the Bogolyubov
coefficients $\alpha$ and $\beta$ in the analysis, 
 (ii) to
 demonstrates a technique of finding out the 
constraint relations between  the complex parameters $\alpha$ and $\beta$ using the theoretical  normalization
condition and observational data of power spectrum and bispectrum from CMB.

It was quite clear right from the beginning that,
because one can never determine 
four parameters from three pieces of information analytically,
one cannot determine the initial vacuum (i.e., the set \{$a_1$, $a_2$, $b_1$, $b_2$\}, each pair representing one Bogolyubov coefficient) conclusively from CMB data alone. 
In this article, 
we did not even try to do that.
What we intended to do is rather to look at how far one can go in this direction using CMB.
We tried to justify that one can at best give an allowed region
of the parameters representing initial vacuum by utilizing the present uncertainties
in the determination of $f_{NL}$, and that the Bunch-Davies and quite a handful of non-Bunch-Davies vacuua
are equally likely so far as latest CMB data is concerned.
In principle, the analysis can also be subject to the next generation surveys with possibly
 reduced error bars for $f_{NL}$ and a
possible detection of primordial non-Gaussinities can be the most optimistic scenario.

The primary limitation of the analysis is self-explanatory.
Due to less number of information than the number of unknowns, we could not determine the individual errors
in the parameters $a_1$, $a_2$, $b_1$, $b_2$.
Individual errors in those parameters
can be obtained only when we have a supplementary piece of information, either from CMB or from 
an altogether different observation.
This one can do by
cross-correlating CMB data with other existing or forthcoming datasets. 
We are already in the process of exploring one such idea \cite{our21} and 
a rigorous numerical analysis in this direction is in order.
We hope
 to address some of these issues in due course.
 
%%%%%%%%%%%%%%%%%%%%%%%%%%%%%%%%%%%%%%%%%%%%%%%%%%%%%%%%%%%%%%%%%%%%%%%%%%%%%%%%%%%%%%%%%%%%%%%%%%%%%%%%%%%%%%%%%%%%%%%%%%%%%%%%%%%%%%%%%%%%%%%%%%%%%%%%%%%%%%%%%%%%%%%%%%%%%%%%%%%%%%%%%%%%%%%%%%%%%%%%%%%%%%%%%%%%%%%%%%%%%
%%%%%%%%%%%%%%%%%%%%%%%%%%%%%%%%%%%%%%%%%%%%%%%%%%%%%%%%%%%%%%%%%%%%%%%%%%%%%%%%%%%%%%%%%%%%%%%%%%%%%%%%%%%%%%%%%%%%%%%%%%%%%%%%%%%%%%%%%%%%%%%%%%%%%%%%%%%%%%%%%%%%%%%%%%%%%%%%%%%%%%%%%%%%%%%%%%%%%%%%

\section*{Acknowledgments}

DC thanks ISI Kolkata for financial support through Senior Research Fellowship. SP
thanks Alexander von Humboldt Foundation, Germany for partial support through
an alumni support grant. We gratefully acknowledge the computational facilities of
ISI Kolkata.
%%%%%%%%%%%%%%%%%%%%%%%%%%%%%%%%%%%%%%%%%%%%%%%%%%%%%%%%%%%%%%%%%%%%%%%%%%%%%%%%%%%%%%%%%%%%%%%%%%%%%%%%%%%%%%%%%%%%%%%%
%%%%%%%%%%%%%%%%%%%%%%%%%%%%%%%%%%%%%%%%%%%%%%%%%%%%%%%%%%%%%%%%%%%%%%%%%%%%%%%%%%%%%%%%%%%%%%%%%%%%%%%%%%%%%%%%%%%%%%%%%%%%%%%%%%%%%%%%%%%%%%%%%%%%%%%%%%%%%%%%%%%%%%%%%%%%%%%%%%%%%%%%%%%%%%%%%%%
\section{Appendix A} \label{appA}

Here we show the detailed calculations of three point correlation function for a generic vacuum.
\\
Contribution from the terms containing $f(\zeta_{n})$ 
\[\langle 0(t)|\zeta_n(t,\textbf{x})\;\zeta_n(t,\textbf{y})\;f(\zeta_n(t,\textbf{z}))
|0(t)\rangle=\;\langle 0|\zeta_n(t,\textbf{x})\;\zeta_n(t,\textbf{y})\;
f(\zeta_n(t,\textbf{z}))|0\rangle\;+...\]

From the equation (\ref{fzeta}) we get
\begin{multline}
\langle 0|\zeta_n\left(t,\textbf{x}\right) \zeta_n\left(t,\textbf{y}\right) f\left(\zeta_n\left(t,\textbf{z}\right)\right)
|0\rangle \\ =\int \frac{d^3\textbf{k}}{\left(2 \pi\right)^3} \frac{d^3 \textbf{m}}{\left(2 \pi \right)^3} 
\frac{d^3\textbf{n}}{\left(2 \pi\right)^3} e^{i\left( \textbf{k}. \textbf{x}+\textbf{m}.\textbf{y}+ \textbf{n}. \textbf{z}\right)} \left(2 \pi\right)^3 \delta^3 \left(\textbf{k}+\textbf{m}+\textbf{n}\right)\qquad \qquad ~~~~~~~~~~~~~\\ ~~~~~~~\qquad  \left\{ \left(\frac{\ddot{\phi}}
{\dot{\phi}\dot{\rho}} + \frac{1}{2} \frac{1}{M^2_{pl}} 
\frac{\dot{\phi}^2}{\dot{\rho}^2}\right) G^>_{k}\left(t,t\right)G^>_{m}\left(t,t\right)+ 
\frac{1}{\dot{\rho}} \left[\dot{G}^>_{k}\left(t,t\right)G^>_{m}\left(t,t\right)+G^>_{k} \left(t,t\right)\dot{G}^>_{m}\left(t,t\right)\right]\right.\\ \left. ~~~~  \qquad~+ \frac{1}{2}\frac{e^{-2\rho}}
{\dot{\rho}^2}(\textbf{k}.\textbf{m})G^>_{k}\left(t,t\right)G^>_{m}\left(t,t \right)- 
\frac{1}{2}\frac{e^{-2\rho}}{\dot{\rho}^2}\frac{\left(k^2+\textbf{k}.\textbf{m}\right)\left(m^2+\textbf{k}.\textbf{m}\right)}{|\textbf{k}+\textbf{m}|^2} G^>_{k}\left(t,t\right)G^>_{m}\left(t,t\right) \right.\\ \left.  \qquad \qquad + \frac{1}{4} \frac{1}{M^2_{pl}} \frac{\dot{\phi}^2}
{\dot{\rho}^3} \left[\frac{\textbf{k}.\textbf{m}}{m^2} G^>_{k}(t,t)\dot{G}^>_{m}\left(t,t\right) +
\frac{\textbf{k}.\textbf{m}}{k^2} \dot{G}^>_{k}\left(t,t\right)G^>_{m}\left(t,t\right)\right]-  \frac{1}{4} \frac{1}{M^2_{pl}} 
\frac{\dot{\phi}^2}{\dot{\rho}^3}\frac{\left(k^2+\textbf{k}.\textbf{m}\right)\left(m^2 +\textbf{k}.\textbf{m}\right)}{|\textbf{k}+\textbf{m}|^2}\right. \qquad \qquad\\ \left.\qquad\qquad\qquad\qquad~~~~~\qquad\qquad~~~~~\left[\frac{1}{m^2} G^>_{k}\left(t,t\right)
\dot{G}^>_{m}\left(t,t\right)+\frac{1}{k^2} \dot{G}^>_{k}\left(t,t\right) G^>_{m}\left(t,t\right)\right]\right\} \qquad \qquad ~~~~~~~
\end{multline}
%\[\langle 0|\;\zeta_n(t,\textbf{x})\; \zeta_n(t,\textbf{y})\; f(\zeta_n(t,\textbf{z}))
 %|0\rangle\qquad \qquad \qquad \qquad \qquad \qquad \qquad \qquad \qquad
 % \qquad \qquad \qquad \qquad \qquad \qquad\]
%\[= \int \frac{d^3\textbf{k}}{(2 \pi)^3} \frac{d^3 \textbf{m}}{(2 \pi)^3}
% e^{i \textbf{k}. (\textbf{x}-\textbf{z})} e^{i\textbf{m}.(\textbf{y}-\textbf{z})} \qquad \qquad \qquad \qquad \qquad\]
%\[\qquad \qquad \qquad \qquad \Bigg\lbrace \Bigg[\frac{\ddot{\phi}}
%{\dot{\phi}\dot{\rho}} + \frac{1}{2} \frac{1}{M^2_{pl}} 
%\frac{\dot{\phi}^2}{\dot{\rho}^2}\Bigg] G^>_{k}(t,t)G^>_{m}(t,t)+ 
%\frac{1}{\dot{\rho}} \Bigg[\dot{G}^>_{k}(t,t)G^>_{m}(t,t)+G^>_{k}(t,t)\dot{G}^>_{m}(t,t)\Bigg]\]
%\[\qquad \qquad \qquad \qquad + \frac{1}{2}\frac{e^{(-2\rho)}}
%{\dot{\rho}^2}(\textbf{k}.\textbf{m})G^>_{k}(t,t)G^>_{m}(t,t)- 
%\frac{1}{2}\frac{e^{(-2\rho)}}{\dot{\rho}^2}\frac{(k^2+\textbf{k}.\textbf{m})(m^2+\textbf{k}.
%\textbf{m})}{|\textbf{k}+\textbf{m}|^2} G^>_{k}(t,t)G^>_{m}(t,t)\]
%\[\qquad  + \frac{1}{4} \frac{1}{M^2_{pl}} \frac{\dot{\phi}^2}
%{\dot{\rho}^3} \Bigg[\frac{\textbf{k}.\textbf{m}}{m^2} G^>_{k}(t,t)\dot{G}^>_{m}(t,t) +
% \frac{\textbf{k}.\textbf{m}}{k^2} \dot{G}^>_{k}(t,t)G^>_{m}(t,t)\Bigg]\]
%\[\qquad \qquad \qquad \qquad - \frac{1}{4} \frac{1}{M^2_{pl}} 
%\frac{\dot{\phi}^2}{\dot{\rho}^3}\frac{(k^2+\textbf{k}.\textbf{m})(m^2
%+\textbf{k}.\textbf{m})}{|\textbf{k}+\textbf{m}|^2}\Bigg[\frac{1}{m^2} G^>_{k}(t,t)
%\dot{G}^>_{m}(t,t)+\frac{1}{k^2} \dot{G}^>_{k}(t,t) G^>_{m}(t,t)\Bigg]\Bigg\rbrace\]
%\\
Now, the contribution from the three point function of the shifted field $\zeta_n$, generates due the presence of interaction Hamiltonian(\ref{int}),
\[\langle 0\left(t\right)|\;\zeta_n\left(t,\textbf{x}\right)\; \zeta_n(t,\textbf{y})\; \zeta_n\left(t,\textbf{z}\right)
 |0\left(t\right)\rangle
  = -i \int^t_{-\infty} \langle 0|\;[\zeta\left(t,\textbf{x}\right) \;\zeta\left(t,\textbf{y}\right) \;\zeta\left(t,\textbf{z}\right),H_I]\;|0\rangle\]

\[H_I\left(t\right) = - \int d^3\textbf{x}\left[ \emptyset_1 \left(t,\textbf{x}\right) + 
\emptyset_2 \left(t,\textbf{x}\right) + \emptyset_3 \left(t,\textbf{x}\right)\right]\]
Where,
\[\emptyset_1 = \frac{1}{4}  e^{3\rho} 
\frac{1}{M^{2}_{pl}}  \frac{\dot{\phi}^4}{\dot{\rho}^4} \dot{\zeta^2_{n}} \zeta_{n} \;\]
\[ \emptyset_2 = \frac{1}{4} e^{\rho} 
\frac{1}{M^{2}_{pl}} \frac{\dot{\phi}^4}{\dot{\rho}^4} \zeta_{n} 
\partial_{k}\zeta_{n}\; \partial^k \zeta_{n}\]
\[ \emptyset_3 = -\frac{1}{2} e^{3\rho} 
\frac{1}{M^{2}_{pl}} \frac{\dot{\phi}^4}{\dot{\rho}^4} \dot{\zeta_{n}}
\partial_{k}\zeta_{n} \partial^k (\partial^{-2} \dot{\zeta_{n}}) \]
\begin{multline}
\langle\emptyset_1(t)\rangle=\frac{i}{2}\frac{1}{M^2_{pl}}
\int_{-\infty}^t dt' e^{3\rho(t')}\frac{\dot{\phi}^4}{\dot{\rho}^4} 
\int \frac{d^3\textbf{k}}{(2 \pi)^3} \frac{d^3 \textbf{m}}{(2 \pi)^3} 
\frac{d^3\textbf{n}}{\left(2 \pi\right)^3} e^{i\left( \textbf{k}. \textbf{x}+\textbf{m}.\textbf{y}+ \textbf{n}. \textbf{z}\right)} (2 \pi)^3 \delta^3 \left(\textbf{k}+\textbf{m}+\textbf{n}\right)\\ \qquad \qquad  \left[\dot{G}^>_{k}\left(t,t'\right)
\dot{G}^>_{m}\left(t,t'\right)G^>_{n}\left(t,t'\right)- \dot{G}^<_{k}\left(t,t'\right)\dot{G}^<_{m}\left(t,t'\right)G^<_{n} \left(t,t'\right) +\dot{G}^>_{k}\left(t,t'\right)G^>_{m}\left(t,t'\right)
\dot{G}^>_{n}\left(t,t'\right)\right. \\ \qquad \qquad  \left.-\dot{G}^<_{k}\left(t,t'\right)G^<_{m}\left(t,t'\right)\dot{G}^<_{n} \left(t,t'\right) +G^>_{k}\left(t,t'\right)\dot{G}^>_{m}\left(t,t'\right)
\dot{G}^>_{n}\left(t,t'\right)- G^<_{k}\left(t,t'\right)\dot{G}^<_{m} \left(t,t'\right)\dot{G}^<_{n}\left(t,t'\right)\right]
\end{multline}
%\[\langle\emptyset_1(t)\rangle=\frac{i}{2}\frac{1}{M^2_{pl}}\int_{-\infty}^t dt' e^{3\rho(t')}\frac{\dot{\phi}^4}{\dot{\rho}^4}\int \frac{d^3\textbf{k}}{(2 \pi)^3} \frac{d^3 \textbf{m}}{(2 \pi)^3}\frac{d^3\textbf{n}}{(2 \pi)^3} e^{(i \textbf{k}. \textbf{x})} e^{(i\textbf{m}.\textbf{y})} e^{(i \textbf{n}. \textbf{z})} (2 \pi)^3 \delta^3(\textbf{k}+\textbf{m}+\textbf{n}) \]
%\[~~~\qquad\qquad \lbrace\dot{G}^>_{k}(t,t')\dot{G}^>_{m}(t,t')G^>_{n}(t,t')-\dot{G}^<_{k}(t,t')\dot{G}^<_{m}(t,t')G^<_{n}(t,t') +\dot{G}^>_{k}(t,t')G^>_{m}(t,t')\dot{G}^>_{n}(t,t')\]
%\[~~~\qquad\qquad- \dot{G}^<_{k}(t,t')G^<_{m}(t,t')\dot{G}^<_{n}(t,t') +G^>_{k}(t,t')\dot{G}^>_{m}(t,t')\dot{G}^>_{n}(t,t')- G^<_{k}(t,t')\dot{G}^<_{m}(t,t')\dot{G}^<_{n}(t,t')\rbrace\]
\begin{multline}
\langle\emptyset_2(t)\rangle=-\frac{i}{2}\frac{1}{M^2_{pl}}\int_{-\infty}^t dt' e^{\rho\left(t'\right)}
\frac{\dot{\phi}^4}{\dot{\rho}^4} \int \frac{d^3\textbf{k}}
{\left(2 \pi\right)^3} \frac{d^3 \textbf{m}}{\left(2 \pi\right)^3} \frac{d^3\textbf{n}}{\left(2 \pi\right)^3} 
e^{i\left( \textbf{k}. \textbf{x}+\textbf{m}.\textbf{y}+ \textbf{n}. \textbf{z}\right)}
\left(2 \pi\right)^3 \delta^3 \left(\textbf{k}+\textbf{m}+\textbf{n}\right)\\
\qquad \qquad \quad  \left(\textbf{k}.
\textbf{m}+\textbf{k}.\textbf{n}+\textbf{m}.\textbf{n}\right) \left[ G^>_{k}\left(t,t'\right)G^>_{m}\left(t,t'\right)G^>_{n}\left(t,t'\right)- 
G^<_{k}\left(t,t'\right)G^<_{m}\left(t,t'\right)G^<_{n}\left(t,t'\right)\right] 
\end{multline}
%\[\langle\emptyset_2(t)\rangle=-\frac{i}{2}\int_{-\infty}^t dt' e^{\rho(t')}\frac{1}{M^2_{pl}}\frac{\dot{\phi}^4}{\dot{\rho}^4} \int \frac{d^3\textbf{k}}{(2 \pi)^3} \frac{d^3 \textbf{m}}{(2 \pi)^3} \frac{d^3\textbf{n}}{(2 \pi)^3}e^{i \textbf{k}. \textbf{x}} e^{i\textbf{m}.\textbf{y}} e^{i \textbf{n}. \textbf{z}} (2 \pi)^3 \delta^3(\textbf{k}+\textbf{m}+\textbf{n}) \]
%\[~  \Bigg\lbrace[\textbf{k}.\textbf{m}+\textbf{k}.\textbf{n}+\textbf{m}.\textbf{n}] [ G^>_{k}(t,t')G^>_{m}(t,t')G^>_{n}(t,t')-G^<_{k}(t,t')G^<_{m}(t,t')G^<_{n}(t,t')]\Bigg\rbrace\]
\begin{multline}
\langle\emptyset_3\left(t\right)\rangle=-\frac{i}{2}\frac{1}{M^2_{pl}}\int_{-\infty}^t dt'
e^{3\rho\left(t'\right)}\frac{\dot{\phi}^4}{\dot{\rho}^4} \int \frac{d^3\textbf{k}}{\left(2 \pi\right)^3} 
\frac{d^3 \textbf{m}}{\left(2 \pi\right)^3} \frac{d^3\textbf{n}}{\left(2 \pi\right)^3} e^{i\left( \textbf{k}. \textbf{x}+\textbf{m}.\textbf{y}+ \textbf{n}. \textbf{z}\right)} \left(2 \pi\right)^3 \delta^3\left(\textbf{k}+\textbf{m}+\textbf{n}\right)\\
~~~~~~\left\{\left(
\frac{\textbf{k}.\textbf{n}}{k^2}+\frac{\textbf{m}.\textbf{n}}{m^2}\right)
\left[\dot{G}^>_{k}(t,t')\dot{G}^>_{m}(t,t')G^>_{n}\left(t,t'\right)- \dot{G}^<_{k}
\left(t,t'\right)\dot{G}^<_{m}\left(t,t'\right)G^<_{n}\left(t,t'\right)\right]\right. \\ \left.
~~~~\qquad +\left( \frac{\textbf{k}.
	\textbf{m}}{k^2}+\frac{\textbf{n}.\textbf{m}}{n^2}\right)\left[\dot{G}^>_{k}\left(t,t'\right)G^>_{m}\left(t,t'\right)
\dot{G}^>_{n}\left(t,t'\right)- \dot{G}^<_{k}\left(t,t'\right)G^<_{m}\left(t,t'\right)\dot{G}^<_{n} \left(t,t'\right)\right] \right. \\ \left.
~~~~\qquad +\left( \frac{\textbf{m}.\textbf{k}}{m^2}
+\frac{\textbf{n}.\textbf{k}}{n^2}\right)\left[G^>_{k}\left(t,t'\right)\dot{G}^>_{m}\left(t,t'\right)
\dot{G}^>_{n}\left(t,t'\right)- G^<_{k}\left(t,t'\right)\dot{G}^<_{m}\left(t,t'\right)\dot{G}^<_{n}(t,t')\right]
\right\}
\end{multline}
%\[\langle\emptyset_3(t)\rangle=-\frac{i}{2}\frac{1}{M^2_{pl}}\int_{-\infty}^t dt' e^{3\rho(t')}\frac{\dot{\phi}^4}{\dot{\rho}^4} \int \frac{d^3\textbf{k}}{(2 \pi)^3} \frac{d^3 \textbf{m}}{(2 \pi)^3} \frac{d^3\textbf{n}}{(2 \pi)^3} e^{(i \textbf{k}. \textbf{x})}  e^{(i\textbf{m}.\textbf{y})} e^{(i \textbf{n}. \textbf{z})} (2 \pi)^3 \delta^3(\textbf{k}+\textbf{m}+\textbf{n}) \]
%\[~ \Bigg\lbrace\left[ \frac{\textbf{k}.\textbf{n}}{k^2}+\frac{\textbf{m}.\textbf{n}}{m^2}\right] [\dot{G}^>_{k}(t,t')\dot{G}^>_{m}(t,t')G^>_{n}(t,t')- \dot{G}^<_{k} (t,t')\dot{G}^<_{m}(t,t')G^<_{n}(t,t')]\]
%\[~+\left[ \frac{\textbf{k}.\textbf{m}}{k^2}+\frac{\textbf{n}.\textbf{m}}{n^2}\right][\dot{G}^>_{k}(t,t')G^>_{m}(t,t')\dot{G}^>_{n}(t,t')- \dot{G}^<_{k}(t,t')G^<_{m}(t,t')\dot{G}^<_{n}(t,t')]\]
%\[~+\left[ \frac{\textbf{m}.\textbf{k}}{m^2}+\frac{\textbf{n}.\textbf{k}}{n^2}\right][G^>_{k}(t,t')\dot{G}^>_{m}(t,t')\dot{G}^>_{n}(t,t')- G^<_{k}(t,t')\dot{G}^<_{m}(t,t')\dot{G}^<_{n}(t,t')]\Bigg\rbrace\]
\underline{\textbf{Propagator or Greens Function :-}}  
\\
\\
Here we will have two propagators,
\\
\\
$G^{>}\left(t,\textbf{x};t',\textbf{y}\right)=\langle 0|\zeta_{n}\left(t,\textbf{x}\right)\zeta_{n}\left(t',\textbf{y}
\right)|0\rangle=\int \frac{d^{3}\textbf{k}}{\left(2\pi\right)^{3}} e^{i\textbf{k}.\left(\textbf{x}-\textbf{y}\right)}
G^{>}_{k}\left(t,t'\right)$~~~~~~ when t>t'\\\\
$G^{<}\left(t,\textbf{x};t',\textbf{y}\right)=\langle 0|\zeta_{n}\left(t',\textbf{y}\right)\zeta_{n}\left(t,\textbf{x}\right)
|0\rangle=\int \frac{d^{3}\textbf{k}}{\left(2\pi\right)^{3}} e^{i\textbf{k}.\left(\textbf{x}-\textbf{y}\right)}
G^{<}_{k}\left(t,t'\right)$~~~~~~ when t<t'\\\\
$|0\rangle$~= Initial vacuum of free theory\\
\\Explicit expressions of the Greens function are as follows,

\[\lim_{t\to\infty} G^{>}_{k}(t,t)~=~\frac{A}{k^3},\qquad \qquad 
\lim_{t \to \infty}\dot{G^{>}_{k}}(t,t)~=~\frac{B}{k^3}, \]

For the limiting condition $k\tau' \to \infty$,
\[~~~~\lim_{k\tau \to 0}  G^{>}_{k}\left(t,t'\right)\;=\;\frac{1}{k^3} (k \tau') 
\left(  C_3 e^{ik \tau'} + C_4 e^{-ik \tau'}\right), ~~ \lim_{k\tau \to 0}\; \dot{G}^{>}_{k}\left(t,t'\right)\;=\;\frac{1}{k^3}  \left (k\tau'\right)^2\left( C_{7} e^{ik\tau'}
 +  C_{8} e^{-ik\tau'}  \right)\]

Where,
\\
\[ A = \frac{H^2\left(4\nu^2-1\right)^2\left(\epsilon -1\right)^2}{64\pi \epsilon M^{2}_{pl}}\left(|\alpha|^{2} 
 + |\beta|^{2} - ie^{i\nu \pi} \alpha^*\beta + ie^{-i\nu \pi} \beta^*\alpha\right),\]

\[ B = -\frac{H^3\left(4\nu^2-1\right)^2\left(\epsilon-1\right)^2}{32\pi  M^{2}_{pl} }\left(|\alpha|^2+
|\beta|^2-i\alpha^* \beta e^{i \nu \pi}+i\beta^* \alpha e^{-i \nu \pi}\right),\]

\[C_3\;=\;\frac{-i H^2\left(4\nu^2-1\right)\left(\epsilon-1\right)^2}{8\pi \epsilon M^{2}_{pl}}
\left(|\alpha|^2-i\alpha^* \beta e^{i \nu \pi}\right),\qquad\qquad\]
\[~ C_4\;=\;\frac{-i H^2\left(4\nu^2-1\right)\left(\epsilon-1\right)^2}{8\pi \epsilon M^{2}_{pl}}
\left(-|\beta|^2-i\alpha \beta^* e^{-i \nu \pi}\right), \quad\qquad\]

\[~~~~~~\quad C_7=\frac{ H^3\left(4\nu^2-1\right)\left(\epsilon-1\right)^3}{8\pi \epsilon M^{2}_{pl}}
\left(|\alpha|^2-i\alpha^* \beta e^{i \nu \pi}\right)~, \qquad~ C_8=\frac{H^3\left(4\nu^2-1\right)\left(\epsilon-1\right)^3}{8\pi \epsilon M^{2}_{pl}}
\left(|\beta|^2+i\alpha \beta^* e^{-i \nu \pi}\right)~~~ \]

We have used these results in Section \ref{sec3} and subsequent sections.

%%%%%%%%%%%%%%%%%%%%%%%%%%%%%%%%%%%%%%%%%%%%%%%%%%%%%%%%%%%%%%%%%%%%%%%%%%%%%%%%%%%%%%%%%%%%%%%%%%%%%%%%%%%%%%%%%%%%%%%%%%%%%%%%%%%%%%%%%%%%%
\section{Appendix B} \label{appB}

This section is dedicated to a detailed derivation of Bispectrum for a generic vacuum.\\
The contributions from $f(\zeta_{n})$ terms,
%\[\langle \emptyset_0(t) \rangle =\langle 0|\zeta_{n}(t,\textbf{x})\;\zeta_{n}(t,\textbf{y})\;f(\zeta_{n}(t,\textbf{z}))|0\rangle +\langle 0|\zeta_{n}(t,\textbf{x})\;f(\zeta_{n}(t,\textbf{y}))\;\zeta_{n}(t,\textbf{z})|0\rangle\; \qquad\]\[\qquad\qquad\qquad\qquad~~~~~\qquad\qquad\qquad\qquad~~~~~+\;\langle 0|f(\zeta_{n}(t,\textbf{x}))\;\zeta_{n}(t,\textbf{y})\;\zeta_{n}(t,\textbf{z})|0\rangle\]
\begin{multline}
\langle \emptyset_0(t) \rangle =\langle 0|\zeta_{n}(t,\textbf{x})\;\zeta_{n}(t,\textbf{y})\;f(\zeta_{n}
(t,\textbf{z}))|0\rangle +\langle 0|\zeta_{n}(t,\textbf{x})\;
f(\zeta_{n}(t,\textbf{y}))\;\zeta_{n}(t,\textbf{z})|0\rangle\\ \qquad\qquad\qquad\qquad\qquad\qquad\qquad\qquad +\langle 0
|f(\zeta_{n}(t,\textbf{x}))\;\zeta_{n}(t,\textbf{y})\;\zeta_{n}(t,\textbf{z})|0\rangle \\
= \frac{H^4}{32 \epsilon^2 M^4_{pl}} \int \frac{d^3 \textbf{k}
}{(2 \pi)^3} \frac{d^3 \textbf{m}}{(2 \pi)^3} \frac{d^3 \textbf{n}}{(2 \pi)^3} e^{i( \textbf{k}. 
	\textbf{x}+\textbf{m}.\textbf{y}+ \textbf{n}. \textbf{z})} 
(2 \pi)^3 \delta^3(\textbf{k}+\textbf{m}+\textbf{n})\frac{1}{k^3 m^3 n^3}\mathcal{A}_0
\end{multline}
where
\begin{multline}
\mathcal{A}_0=\frac{32 \epsilon^2 M^4_{pl}}{H^4}\left\{ \left[ \left(\frac{\ddot{\phi}}{\dot{\phi}\dot{\rho}}+ 
\frac{1}{2}\frac{1}{M^2_{pl}}\;\frac{\dot{\phi}^2}{\dot{\rho}^2}\right) \left(k^3+m^3+n^3\right)\right] A^2+ \frac{2}{\dot{\rho}}\left(k^3+m^3+n^3\right)AB  \right. 
\\
\qquad \quad  \left. +\frac{1}{4} \frac{1}{M^{2}_{pl}} \frac{\dot{\phi}^2}{\dot{\rho}^3} 
\frac{1}{k^2 m^2 n^2} \left[ \left(k^2+m^2\right)\left(\textbf{k}.\textbf{m}\right)n^5 + \left(k^2+n^2\right)\left(\textbf{k}.\textbf{n}\right)m^5 + \left(m^2+n^2\right)\left(\textbf{m}.\textbf{n}\right)k^5 \right]AB \right.  
\\
  \qquad \quad  \left. - \frac{1}{4} \frac{1}{M^{2}_{pl}} \frac{\dot{\phi}^2}{\dot{\rho}^3} \frac{1}{k^2 m^2 n^2}\left[ \frac{\left(k^2 + \textbf{k}.\textbf{m}\right)\left(m^2 + \textbf{k}.\textbf{m}\right)}{|\textbf{k} + \textbf{m}|^{2}} \left(k^2+m^2\right)n^5+\frac{\left(k^2 + \textbf{k}.\textbf{n}\right)\left(n^2 + \textbf{k}.\textbf{n}\right)}{|\textbf{k} + \textbf{n}|^{2}} \left(k^2+n^2\right)m^5 \right. \right. \\ 
\qquad \qquad  \left. \left. + \frac{\left(m^2 + \textbf{m}.\textbf{n}\right)\left(n^2 + \textbf{m}.\textbf{n}\right)}{|\textbf{m} + \textbf{n}|^{2}}\left(m^2+n^2\right)k^5 \right] AB \right\}
\end{multline}

Contribution from the terms evaluated using interaction Hamiltonian,
\begin{equation}
\langle \emptyset_1(t) \rangle = \frac{H^4}{32 \epsilon^2 M^4_{pl}} \int   \frac{d^3 \textbf{k}
}{(2 \pi)^3} \frac{d^3 \textbf{m}}{(2 \pi)^3} \frac{d^3 \textbf{n}}{(2 \pi)^3} e^{i( \textbf{k}. 
	\textbf{x}+\textbf{m}.\textbf{y}+ \textbf{n}. \textbf{z})} 
(2 \pi)^3 \delta^3(\textbf{k}+\textbf{m}+\textbf{n}) \frac{1}{k^3 m^3 n^3} \mathcal{A}_1
\end{equation}
where
\begin{multline}
\mathcal{A}_1=\frac{i 128 \epsilon^4 M^6_{pl} }{(\epsilon - 1)^4 H^8 }\left\{  D_1 \frac{kmn(km+kn+mn)}{(k+m+n)^2} +
D_2 \left[ \frac{k^2m^2n}{(k+m-n)^2} + \frac{k^2mn^2}{(k+n-m)^2} \right. \right.  \\  \left.\left. +\frac{km^2n^2}{(m+n-k)^2} \right] + \left(D_3+D_4\right) \left[ \frac{k^2m^2n}{(m+n-k)^2} + \frac{k^2m^2n}{(k+n-m)^2}+ \frac{k^2mn^2}{(m+n-k)^2} \right. \right.  \\ \left. \left.  + \frac{k^2mn^{2}}{(k+m-n)^2} +\frac{km^2n^2}{(k+m-n)^2} +\frac{km^2n^2}{(k+n-m)^2} \right]\right\}
\end{multline}
\begin{equation}
\langle \emptyset_2(t) \rangle =\frac{H^4}{32 \epsilon^2 M^4_{pl}} \int \frac{d^3 \textbf{k}
}{(2 \pi)^3} \frac{d^3 \textbf{m}}{(2 \pi)^3} \frac{d^3 \textbf{n}}{(2 \pi)^3} e^{i( \textbf{k}.\textbf{x}+\textbf{m}.\textbf{y}+ \textbf{n}. \textbf{z})}(2 \pi)^3 \delta^3(\textbf{k}+\textbf{m}+\textbf{n}) \frac{1}{k^3 m^3 n^3} \mathcal{A}_2
\end{equation}
where
\begin{multline}
\mathcal{A}_2=- \frac{i 128 \epsilon^4 M^6_{pl}}{  (\epsilon - 1)^2 H^6}\left( \textbf{k}.\textbf{m}+
 \textbf{k}.\textbf{n}+\textbf{m}.\textbf{n} \right) 
\left\{\left(C^3_3+C^3_4\right)\frac{kmn}
{(k+m+n)^2} + \left(C^2_3C_4+C^2_4C_3\right)\right. \\ \left.  \left[ \frac{kmn}{(-k+m+n)^2}  +
 \frac{kmn}{(k-m+n)^2} +\frac{kmn}{(k+m-n)^2} \right] \right\}
\end{multline}

%\begin{multline}
%\langle \emptyset_2(t) \rangle = - \frac{i 128 \epsilon^4 M^6_{pl}}{  (\epsilon - 1)^2 H^6} \int \frac{d^3 \textbf{k}
%}{(2 \pi)^3} \frac{d^3 \textbf{m}}{(2 \pi)^3} \frac{d^3 \textbf{n}}{(2 \pi)^3} e^{i( \textbf{k}.\textbf{x}+\textbf{m}.\textbf{y}+ \textbf{n}. \textbf{z})}(2 \pi)^3 \delta^3(\textbf{k}+\textbf{m}+\textbf{n}) \frac{1}{k^3 m^3 n^3} \frac{H^4}{32 \epsilon^2 M^4_{pl}} 
%\\
%\left( \textbf{k}.\textbf{m}+
 %\textbf{k}.\textbf{n}+\textbf{m}.\textbf{n} \right) 
%\left\{\left(C^3_3+C^3_4\right)\frac{kmn}
%{(k+m+n)^2} + \left(C^2_3C_4+C^2_4C_3\right)\right. \\ \left.  \left[ \frac{kmn}{(-k+m+n)^2}  +
% \frac{kmn}{(k-m+n)^2} +\frac{kmn}{(k+m-n)^2} \right] \right\} 
%\end{multline}
\begin{equation}
\langle \emptyset_3(t)\rangle  = \frac{H^4}{32 \epsilon^2 M^4_{pl}} \int \frac{d^3 \textbf{k}
}{(2 \pi)^3} \frac{d^3 \textbf{m}}{(2 \pi)^3} \frac{d^3 \textbf{n}}{(2 \pi)^3} e^{i( \textbf{k}. 	\textbf{x}+\textbf{m}.\textbf{y}+ \textbf{n}. \textbf{z})} (2 \pi)^3 \delta^3(\textbf{k}+\textbf{m}+\textbf{n})\frac{1}{k^3 m^3 n^3}\mathcal{A}_3
\end{equation}
where
\begin{multline}
\mathcal{A}_3= - \frac{i 128 \epsilon^4 M^6_{pl}}{H^8(\epsilon - 1)^4 }\left\{  D_1 \left[\frac{m^2(\textbf{k}.\textbf{n})+k^2(\textbf{m}.\textbf{n})}{km}\frac{kmn}{(k+m+n)^2}+ \frac{n^2(\textbf{k}.\textbf{m}) +k^2(\textbf{n}.\textbf{m})}{kn}\frac{kmn}{(k+m+n)^2} \right. \right. 
 \\ \left. \left. +\frac{n^2(\textbf{m}.\textbf{k}) +m^2(\textbf{n}.\textbf{k})}{mn}\frac{kmn}{(k+m+n)^2} \right] +D_2\left[\frac{m^2(\textbf{k}.\textbf{n})+k^2(\textbf{m}.\textbf{n})}{km}\frac{kmn}{(k+m-n)^2} \right. \right.
 \\  \left. \left. + \frac{n^2(\textbf{k}.\textbf{m})+k^2(\textbf{n}.\textbf{m})}{kn}\frac{kmn}{(k+n-m)^2}  +\frac{n^2(\textbf{m}.\textbf{k})+m^2(\textbf{n}.\textbf{k})}{mn}\frac{kmn}{(m+n-k)^2} \right] \right. \qquad 
 \\ \qquad \quad \left. +(D_3+D_4)\left[\frac{m^2(\textbf{k}.\textbf{n})+k^2(\textbf{m}.\textbf{n})}{km}
\left(\frac{kmn}{(m+n-k)^2}+\frac{kmn}{(k+n-m)^2}\right)+\frac{n^2(\textbf{k}.\textbf{m})+k^2(\textbf{n}.\textbf{m})}{kn} \right. \right. 
\\ \qquad \quad \left. \left. \left(\frac{kmn}{(m+n-k)^2}+\frac{kmn}{(k+m-n)^2}\right) + \frac{n^2(\textbf{m}.\textbf{k})+m^2(\textbf{n}.\textbf{k})}
{mn}\left(\frac{kmn}{(k+m-n)^2}+\frac{kmn}{(k+n-m)^2}\right)\right] \right\}
\end{multline}
Where,
\[D_1 = C_3C_7^2+C_4C_8^2,\qquad D_2 = C_3C_8^2+C_4C_7^2,\qquad 
D_3 = C_3C_7C_8,\qquad D_4 = C_4C_7C_8 \]

\paragraph{Final form of the Bispectrum}

\[\frac{H^4}{32 \epsilon^2 M^4_{pl}} \int\frac{d^3 \textbf{k}
}{(2 \pi)^3} \frac{d^3 \textbf{m}}{(2 \pi)^3} \frac{d^3 \textbf{n}}{(2 \pi)^3} e^{i( \textbf{k}. 
	\textbf{x}+\textbf{m}.\textbf{y}+ \textbf{n}. \textbf{z})} 
(2 \pi)^3 \delta^3(\textbf{k}+\textbf{m}+\textbf{n}) \frac{1}{k^3 m^3 n^3} \left(\mathcal{A}_0 + \mathcal{A}_1 + \mathcal{A}_2 + \mathcal{A}_3 \right)\]
\[ \qquad\qquad\qquad\qquad\qquad\qquad=\frac{H^4}{32 \epsilon^2 M^4_{pl}}\int \frac{d^3 \textbf{k}
}{(2 \pi)^3} \frac{d^3 \textbf{m}}{(2 \pi)^3} \frac{d^3 \textbf{n}}{(2 \pi)^3} e^{i( \textbf{k}. 
	\textbf{x}+\textbf{m}.\textbf{y}+ \textbf{n}. \textbf{z})} 
(2 \pi)^3 \delta^3(\textbf{k}+\textbf{m}+\textbf{n}) \frac{1}{k^3 m^3 n^3}
   \mathcal{A} \qquad \qquad \qquad \qquad \qquad\]
\
where,
$\mathcal{A}$ is defined as\[\mathcal{A} =\mathcal{A}_0 + \mathcal{A}_1 + \mathcal{A}_2 + \mathcal{A}_3\] .\\
Finally, expression of Bispectrum,\[B_{\zeta}
(\textbf{k},\textbf{m},\textbf{n}) = (2 \pi)^3 \delta(
\textbf{k}+\textbf{m}+\textbf{n}) \frac{1}{k^3 m^3 n^3}
 \frac{H^4}{32 \epsilon^2 M^4_{pl}}\mathcal{A} \]
We have used this result to evaluate $f_{NL}$ in section \ref{sec4}.

%%%%%%%%%%%%%%%%%%%%%%%%%%%%%%%%%%%%%%%%%%%%%%%%%%%%%%%%%%%%%%%%%%%%%%%%%%%%%%%%%%%%%%%%%%%%%%%%%%%%%%%%%%%%%%%%%%%%
%%%%%%%%%%%%%%%%%%%%%%%%%%%%%%%%%%%%%%%%%%%%%%%%%%%%%%%%%%%%%%%%%%%%%%%%%%%%%%%%%%%%%%%%%%%%%%%%%%%%%%%%%%%%%%%%%%%%%%%%%%%%%%%%%%%%%%%%

\end{document}